\documentclass[fleqn,11pt]{article}

\usepackage{latexsym,ifthen,epsfig,color}
\usepackage{amsmath,amssymb,amsthm}
\usepackage{float}

\newcommand{\join}{\text{\textcircled{{\footnotesize 1}}}}
\newcommand{\cojoin}{\text{\textcircled{{\footnotesize 0}}}}

\newcommand{\NP}{\ensuremath{\mathbb{NP}}}

\newtheorem{theorem}{Theorem}
\newtheorem{lemma}{Lemma}
\newtheorem{remark}{Remark}
\newtheorem{corollary}{Corollary}
\newtheorem{proposition}{Proposition}

\newtheorem{clai}{Claim}
\newtheorem{observation}{Observation}
\newtheorem{algo}{Algorithm}[section]
\newtheorem{proc}{Procedure}[section]

\begin{document}

\title{Finding Dominating Induced Matchings in $S_{2,2,3}$-Free Graphs in Polynomial Time}

\author{
Andreas Brandst\"adt\footnote{Institut f\"ur Informatik,
Universit\"at Rostock, A.-Einstein-Str.\ 22, D-18051 Rostock, Germany,
{\texttt andreas.brandstaedt@uni-rostock.de}}
\and
Raffaele Mosca\footnote{Dipartimento di Economia, Universit\'a degli Studi ``G.\ D'Annunzio''
Pescara 65121, Italy.
{\texttt r.mosca@unich.it}}
}

\maketitle

\begin{abstract}
Let $G=(V,E)$ be a finite undirected graph. An edge set $E' \subseteq E$ is a {\em dominating induced matching} ({\em d.i.m.}) in $G$ if every edge in $E$ is intersected by exactly one edge of $E'$.
The \emph{Dominating Induced Matching} (\emph{DIM}) problem asks for the existence of a d.i.m.\ in $G$; this problem is also known as the \emph{Efficient Edge Domination} problem; it is the Efficient Domination problem for line graphs.

The DIM problem is \NP-complete even for very restricted graph classes such as planar bipartite graphs with maximum degree 3 and is solvable in linear time for $P_7$-free graphs, and in polynomial time for $S_{1,2,4}$-free graphs as well as for $S_{2,2,2}$-free graphs. In this paper, combining two distinct approaches, we solve it in polynomial time for $S_{2,2,3}$-free graphs.
\end{abstract}

\noindent{\small\textbf{Keywords}:
dominating induced matching;
efficient edge domination;
$S_{2,2,3}$-free graphs;
polynomial time algorithm;
}

\section{Introduction}\label{sec:intro}

Let $G=(V,E)$ be a finite undirected graph. A vertex $v \in V$ {\em dominates} itself and its neighbors. A vertex subset $D \subseteq V$ is an {\em efficient dominating set} ({\em e.d.s.} for short) of $G$ if every vertex of $G$ is dominated by exactly one vertex in $D$.
The notion of efficient domination was introduced by Biggs \cite{Biggs1973} under the name {\em perfect code}.
The {\sc Efficient Domination} (ED) problem asks for the existence of an e.d.s.\ in a given graph $G$ (note that not every graph has an e.d.s.)

A set $M$ of edges in a graph $G$ is an \emph{efficient edge dominating set} (\emph{e.e.d.s.} for short) of $G$ if and only if it is an e.d.s.\ in its line graph $L(G)$. The {\sc Efficient Edge Domination} (EED) problem asks for the existence of an e.e.d.s.\ in a given graph $G$. Thus, the EED problem for a graph $G$ corresponds to the ED problem for its line graph $L(G)$. Note that not every graph has an e.e.d.s. An efficient edge dominating set is also called \emph{dominating induced matching} ({\em d.i.m.} for short)--recall that an edge subset $E' \subseteq E$ is a dominating induced matching in $G$ if every edge in $E$ is intersected by exactly one edge of $E'$.

The EED problem is motivated by applications such as parallel resource allocation of parallel processing systems, encoding theory and network routing---see e.g.\ \cite{GriSlaSheHol1993,HerLozRieZamdeW2015}.
The EED problem is called the {\sc Dominating Induced Matching} (DIM) problem in various papers (see e.g.\ \cite{BraHunNev2010,BraMos2014,CarKorLoz2011,HerLozRieZamdeW2015,KorLozPur2014}); subsequently, we will use this terminology in our manuscript.

In \cite{GriSlaSheHol1993}, it was shown that the DIM problem is \NP-complete; see e.g.\ \cite{BraHunNev2010,CarKorLoz2011,LuKoTan2002,LuTan1998}.
However, for various graph classes, DIM is solvable in polynomial time. For mentioning some examples, we need the following notions:

Let $P_k$ denote the path with $k$ vertices, say $a_1,\ldots,a_k$, and $k-1$ edges $a_ia_{i+1}$, $1 \le i \le k-1$.
Such a path will also be denoted by $(a_1,\ldots,a_k)$. When speaking about a $P_k$ in a graph $G$, we
will always assume that the path is chordless, or, equivalently, that it is an induced subgraph of $G$.

For indices $i,j,k \ge 0$, let $S_{i,j,k}$ denote the graph with vertices $u,x_1,\ldots,x_i$, $y_1,\ldots,y_j$, $z_1,\ldots,z_k$ such that the subgraph induced by $u,x_1,\ldots,x_i$ forms a $P_{i+1}$ $(u,x_1,\ldots,x_i)$, the subgraph induced by $u,y_1,\ldots,y_j$ forms a $P_{j+1}$ $(u,y_1,\ldots,y_j)$, the subgraph induced by $u,z_1,\ldots,z_k$ forms a $P_{k+1}$ $(u,z_1,\ldots,z_k)$, and there are no other edges in $S_{i,j,k}$. Vertex $u$ is called the {\em center} of this $S_{i,j,k}$.
Thus, {\em claw} is $S_{1,1,1}$, and $P_k$ is isomorphic to $S_{0,0,k-1}$.

\medskip

In \cite{HerLozRieZamdeW2015}, it is conjectured that for every fixed $i,j,k$, DIM is solvable in polynomial time for $S_{i,j,k}$-free graphs (actually, an even stronger conjecture is mentioned in \cite{HerLozRieZamdeW2015}); this includes $P_k$-free graphs for $k \ge 8$.
The following results are known:

\begin{theorem}\label{DIMpolresults}
DIM is solvable in polynomial time for
\begin{itemize}
\item[$(i)$]  $S_{1,1,1}$-free graphs $\cite{CarKorLoz2011}$,
\item[$(ii)$] $S_{1,2,3}$-free graphs $\cite{KorLozPur2014}$,
\item[$(iii)$] $S_{2,2,2}$-free graphs $\cite{HerLozRieZamdeW2015}$,
\item[$(iv)$] $S_{1,2,4}$-free graphs $\cite{BraMos2017/2}$,
\item[$(v)$] $S_{1,1,5}$-free graphs $\cite{BraMos2019/1}$,
\item[$(vi)$] $P_7$-free graphs $\cite{BraMos2014}$ $($in this case even in linear time$)$,
\item[$(vii)$] $P_8$-free graphs $\cite{BraMos2017}$, and
\item[$(viii)$] $P_9$-free graphs $\cite{BraMos2019/2}$.
\end{itemize}
\end{theorem}

Based on the two distinct approaches described in \cite{BraMos2017} and in \cite{HerLozRieZamdeW2015,KorLozPur2014} (and combining them as in \cite{BraMos2017/2} and \cite{BraMos2019/1}), we show in this paper that DIM can be solved in polynomial time for $S_{2,2,3}$-free graphs.

\section{Definitions and basic properties}\label{sec:basicnotionsresults}

\subsection{Basic notions}\label{subsec:basicnotions}

Let $G$ be a finite undirected graph without loops and multiple edges. Let $V$ denote its vertex set and $E$ its edge set; let $n=|V|$ and $m=|E|$.
For $v \in V$, let $N(v):=\{u \in V: uv \in E\}$ denote the {\em open neighborhood of $v$}, and let $N[v]:=N(v) \cup \{v\}$ denote the {\em closed neighborhood of $v$}. The {\em degree} of $v$ in $G$ is $d_G(v)=|N(v)|$.
If $xy \in E$, we also say that $x$ and $y$ {\em see each other}, and if $xy \not\in E$, we say that $x$ and $y$ {\em miss each other}. A vertex set $S$ is {\em independent} in $G$ if for every pair of vertices $x,y \in S$, $xy \not\in E$.
A vertex set $Q$ is a {\em clique} in $G$ if every two distinct vertices in $Q$ are adjacent.

For $uv \in E$ let $N(uv):= (N(u) \cup N(v)) \setminus \{u,v\}$ and $N[uv]:= N[u] \cup N[v]$.

For $U \subseteq V$, let $G[U]$ denote the subgraph of $G$ induced by vertex subset $U$. Clearly $xy \in E$ is an edge in $G[U]$ exactly when $x \in U$ and $y \in U$; thus, $G[U]$ will be often denoted simply by $U$ when that is clear in the context.

For $A \subseteq V$ and $B \subseteq V$, $A \cap B = \emptyset$, we say that $A \cojoin B$ ($A$ and $B$ {\em miss each other}) if there is no edge between $A$ and $B$, and $A$ and $B$ {\em see each other} if there is at least one edge between $A$ and $B$. If a vertex $u \notin B$ has a neighbor in $B$ then {\em $u$ contacts $B$}. If every vertex in $A$ sees every vertex in $B$, we denote it by $A \join B$. For $A=\{a\}$, we simply denote  $A \join B$ by $a \join B$, and correspondingly for $A \cojoin B$ by $a \cojoin B$.
If for $A' \subseteq A$, $A' \cojoin (A \setminus A')$, we say that $A'$ is {\em isolated} in $G[A]$.
For graphs $H_1$, $H_2$ with disjoint vertex sets, $H_1+H_2$ denotes the disjoint union of $H_1$, $H_2$, and for $k \ge 2$, $kH$ denotes the disjoint union of $k$ copies of $H$. For example, $2P_2$ is the disjoint union of two edges.

As already mentioned, a {\em path} $P_k$ ({\em cycle} $C_k$, respectively) has $k$ vertices, say $v_1,\ldots,v_k$, and edges $v_iv_{i+1}$, $1 \le i \le k-1$ (and $v_kv_1$, respectively). We say that such a path has length $k-1$ and such a cycle has length $k$. A $C_3$ is called a {\em triangle}.
Let $K_i$, $i \ge 1$, denote the complete graph with $i$ vertices. Clearly, $K_3=C_3$. Let $K_4-e$ or {\em diamond} be the graph with four vertices, say $v_1,v_2,v_3,u$, such that $(v_1,v_2,v_3)$ forms a $P_3$ and $u \join \{v_1,v_2,v_3\}$; its {\em mid-edge} is the edge $uv_2$.
A {\em gem} has five vertices, say $v_1,v_2,v_3,v_4,u$, such that $(v_1,v_2,v_3,v_4)$ forms a $P_4$ and $u \join \{v_1,v_2,v_3,v_4\}$.
A {\em butterfly} has five vertices, say, $v_1,v_2,v_3,v_4,u$, such that $v_1,v_2,v_3,v_4$ induce a $2P_2$ with edges $v_1v_2$ and $v_3v_4$ (the {\em peripheral edges} of the butterfly), and $u \join \{v_1,v_2,v_3,v_4\}$. Vertices $a,b,c,d$ induce a {\em paw} if $b,c,d$ induce a $C_3$ and $a$ has exactly one neighbor in $b,c,d$, say $b$ (then $ab$ is the {\em leaf edge} of the paw).

For a set ${\cal F}$ of graphs, a graph $G$ is called {\em ${\cal F}$-free} if $G$ contains no induced subgraph from ${\cal F}$. If ${\cal F}=\{H\}$ then instead of $\{H\}$-free, $G$ is called $H$-free. If for instance, $G$ is diamond-free and butterfly-free, we say that $G$ is (diamond,butterfly)-free.

\medskip

We often consider an edge $e = uv$ to be a set of two vertices; then it makes sense to say, for example, $u \in e$ and $e \cap e'$ for an edge $e'$.

For two vertices $x,y \in V$, let $dist_G(x,y)$ denote the {\em distance between $x$ and $y$ in $G$}, i.e., the length of a shortest path between $x$ and $y$ in $G$.
The {\em distance between a vertex $z$ and an edge $xy$} is the length of a shortest path between $z$ and $x,y$, i.e., $dist_G(z,xy)= \min\{dist_G(z,v) \mid v \in \{x,y\}\}$.
The {\em distance between two edges} $e,e' \in E$ is the length of a shortest path between $e$ and $e'$, i.e., $dist_G(e,e')= \min\{dist_G(u,v) \mid u \in e, v \in e'\}$.
In particular, this means that $dist_G(e,e')=0$ if and only if $e \cap e' \neq \emptyset$. Obviously, if $M$ is a d.i.m.\ then for every pair $e,e' \in M$, $e \neq e'$, $dist_G(e,e') \ge 2$ holds. A set of edges whose elements have pairwise distance at least 2 is also called {\em induced matching}.

For an edge $xy$, let $N_i(xy)$, $i \ge 1$, denote the {\em distance levels of $xy$}:
$$N_i(xy):=\{z \in V \mid dist_G(z,xy) = i\}.$$

\medskip

Clearly, for a d.i.m.\ $M$ of $G$ and its vertex set $V(M)$, $I:=V \setminus V(M)$ is an independent set in $G$, i.e.,
\begin{equation}\label{IV(M)partition}
V \mbox{ has partition } V = I \cup V(M).
\end{equation}

From now on, let us color all vertices in $I$ white and all vertices in $V(M)$ black.
According to \cite{HerLozRieZamdeW2015}, we also use the following notions:
In the process of finding $I$ and $M$ in $G$, we will assign either color black or color white to the vertices of $G$, and the assignment of one of the two colors to each vertex of $G$ is called a {\em complete coloring} of $G$. If not all vertices of $G$ have been assigned a color, the coloring is said to be {\em partial}.

A partial black-white coloring of $V(G)$ is {\em feasible} if the set of white vertices is an independent set in $G$ and every black vertex has at most one black neighbor. A complete black-white coloring of $V(G)$ is {\em feasible} if the set of white vertices is an independent set in $G$ and every black vertex has exactly one black neighbor. Clearly, $M$ is a d.i.m.\ of $G$ if and only if the black vertices $V(M)$ and the white vertices $V \setminus V(M)$ form a complete feasible coloring of $V(G)$, and a black-white coloring is {\em not feasible} if there is a {\em contradiction}, e.g., an edge with two white vertices etc. (From now on, we do not always mention ``feasible'' when discussing black-white colorings.)

\subsection{Reduction steps, forbidden subgraphs, forced edges, and excluded edges}\label{exclforced}

\begin{observation}[\cite{BraHunNev2010,BraMos2014}]\label{dimC3C5C7C4}
Let $M$ be a dominating induced matching in $G$. Then:
\begin{itemize}
\item[$(i)$] $M$ contains at least one edge of every odd cycle $C_{2k+1}$ in $G$, $k \ge 1$,
and exactly one edge of every odd cycle $C_3$, $C_5$, $C_7$ of $G$.
\item[$(ii)$] No edge of any $C_4$ is in $M$.
\end{itemize}
\end{observation}

(See e.g.\ Observation 2 in \cite{BraMos2014}.)

\medskip

By Observation \ref{dimC3C5C7C4} $(i)$, every $C_3$ contains exactly one $M$-edge. Then, since the pairwise distance of edges in any d.i.m.\ $M$ is at least 2, we have:
\begin{corollary}\label{cly:k4gemfree}
If a graph has a d.i.m.\ then it is $K_4$-free.
\end{corollary}

{\sc Assumption 1.} By Corollary~\ref{cly:k4gemfree}, assume that the input graph is $K_4$-free. For every subset of four vertices, one can check if they induce a $K_4$ (which can be done in polynomial time).

\medskip

If an edge $e \in E$ is contained in {\bf every} d.i.m.\ of $G$, we call it a {\em forced} edge of $G$. If an edge $e \in E$ is {\bf not} contained in {\bf any} d.i.m.\ of $G$, we call it an {\em excluded} edge of $G$ (we can denote this by a red edge-color).

Note that in a graph with d.i.m.\ $M$, the set $M' \subseteq M$ of forced edges is an induced matching. In our final algorithm solving the DIM problem on $S_{2,2,3}$-free graphs, the set $M'$ of forced edges will be computed and for example, it has to be checked whether $M'$ is really an induced matching.

\begin{observation}\label{obs:diamondbutterfly}
The mid-edge of any diamond in $G$ and the two peripheral edges of any induced butterfly are forced edges of $G$.
\end{observation}

In \cite{BraMos2014}, we used forced edges for reducing the graph $G$ to a smaller graph $G'$ such that $G$ has a d.i.m.\ if and only if $G'$ has a d.i.m. Here we combine the reduction approach with the one of \cite{HerLozRieZamdeW2015} as follows. A vertex $v$ is {\em forced to be white} if for every d.i.m.\ $M$ of $G$,
$v \in V \setminus V(M)$. Analogously, a vertex $v$ is {\em forced to be black} if for every d.i.m.\ $M$ of $G$,
$v \in V(M)$. Clearly, if $uv \in E$ and if $u$ and $v$ are forced to be black, then $uv$ is contained in every (possible) d.i.m.\ of $G$.

\medskip

For the correctness of the reduction steps, we have to argue that $G$ has a d.i.m.\ if and only if the reduced graph $G'$ has one (provided that no contradiction arises in the vertex coloring, i.e., it is feasible).

\medskip

Then let us introduce three reduction steps which will be applied later.

\medskip

{\bf Vertex Reduction.} Let $u \in V(G)$. If $u$ is forced to be white, then
\begin{itemize}
\item[$(i)$] color black all neighbors of $u$, and
\item[$(ii)$] remove $u$ from $G$.
\end{itemize}

Let $G'$ be the reduced graph. Clearly, Vertex Reduction is correct, i.e., $G$ has a d.i.m.\ if and only if $G'$ has a d.i.m.

\medskip

{\bf Edge Reduction.} Let $uw \in E(G)$. If $u$ and $w$ are forced to be black, then
\begin{itemize}
\item[$(i)$] color white all neighbors of $u$ and of $w$ (other than $u$ and $w$), and
\item[$(ii)$] remove $u$ and $w$ from $G$.
\end{itemize}

Again, clearly, Edge Reduction is correct, i.e., $G$ has a d.i.m.\ if and only if $G'$ has a d.i.m.

\medskip

{\sc Assumption 2.} By Observation \ref{obs:diamondbutterfly}, assume that the input graph is (diamond,butterfly)-free: One can apply the
Edge Reduction to each mid-edge of any induced diamond, and to each peripheral edge of any induced butterfly (which can be done in polynomial time).

\medskip

The third reduction step is based on vertex degrees. A triangle in $G$ with vertices, say $a,b,c$, is called a {\em peripheral triangle} if $d_G(b)=d_G(c)=2$ (i.e., $N(b)=\{a,c\}$, $N(c)=\{a,b\}$) and $d_G(a) = 3$ (i.e., $a$ has exactly one further neighbor apart from $b,c$).

\medskip

{\bf Peripheral Triangle Reduction.} Let $abc$ be a peripheral triangle of $G$. Then remove $a,b,c$ from $G$.

\medskip

By Observation \ref{dimC3C5C7C4} $(i)$ with respect to $C_3$ and the distance property, we have the following:

\begin{observation}\label{pawleafedge}
If $a,b,c,d$ induce a paw with $C_3$ $bcd$, and $a$ is adjacent to $b$ then its leaf edge $ab$ is excluded.
\end{observation}

\begin{observation}\label{pertriredcorrect}
$G$ has a d.i.m.\ if and only if, by applying Peripheral Triangle Reduction, the reduced graph $G'$ has a d.i.m.
\end{observation}

{\bf Proof.} Let $a,b,c$ induce a peripheral triangle in $G$, and let $u$ be the third neighbor of $a$.

First assume that $M$ is a d.i.m.\ of $G$. Recall that the triangle $a,b,c$ has exactly one $M$-edge. If $ab \in M$ then clearly, $G'$ has a d.i.m.\ $M':=M \setminus \{ab\}$, and similarly if $ac \in M$. Moreover, if $bc \in M$ then clearly, $G'$ has a d.i.m.\ $M':=M \setminus \{bc\}$.

Now assume that $G'$ has a d.i.m.\ $M'$. If $u$ is white then in $G$, $a$ is black and thus, either $M=M' \cup \{ab\}$ or $M=M' \cup \{ac\}$ is a d.i.m.\ of $G$.
If $u$ is black then $M=M' \cup \{bc\}$ is a d.i.m.\ of $G$. 
\qed

\medskip

Let $G_1$ denote the graph with $V(G_1)=\{x_1,\ldots,x_5,y_1,z_1,y_3,z_3\}$ such that $x_1,\ldots,x_5$ induce a $C_5$, $x_1,y_1,z_1$ induce a triangle,
 $x_3,y_3,z_3$ induce a triangle, and there are no other edges in $G_1$.

\medskip

By Observation \ref{dimC3C5C7C4} $(i)$, with respect to $C_3$ and $C_5$, we have:

\begin{observation}\label{obs:G1forcededges}
Every induced $G_1$ contains three forced edges of $G$, namely $y_1z_1$, $y_3z_3$, and $x_4x_5$.
\end{observation}

{\sc Assumption 3.} By Observation \ref{obs:G1forcededges}, assume that the input graph is $G_1$-free: One can apply the
Edge Reduction to the corresponding three forced edges of any induced $G_1$ (which can be done in polynomial time).

\medskip

Let $G_2$ denote the graph with $V(G_2)=\{x_1,\ldots,x_5,y_1,z_1,x^*\}$ such that $x_1,\ldots,x_5$ induce a $C_5$, $x_1,y_1,z_1$ induce a triangle,
 $x_2,x_3,x^*$ induce a triangle, and there are no other edges in $G_2$.

\medskip

By Observation \ref{dimC3C5C7C4} $(i)$, with respect to $C_3$ and $C_5$, and by Observation \ref{pawleafedge}, we have:

\begin{observation}\label{obs:G2forcededge}
Every induced $G_2$ contains a forced edge of $G$, namely $y_1z_1$.
\end{observation}

{\sc Assumption 4.} By Observation \ref{obs:G2forcededge}, assume that the input graph is $G_2$-free: One can apply the
Edge Reduction to the corresponding forced edge of any induced $G_2$ (which can be done in polynomial time).

\medskip

Let $G_3$ denote the graph with $V(G_3)=\{x_1,\ldots,x_5,y,x^*\}$ such that $x_1,\ldots,x_5$ induce a $C_5$, $x_1,x_2,x_3,y$ induce a $C_4$,
 $x_4,x_5,x^*$ induce a triangle, and there are no other edges in $G_3$.

\medskip

By Observation \ref{dimC3C5C7C4} $(i)$, with respect to $C_3$ and $C_5$, by Observation \ref{dimC3C5C7C4} $(ii)$ with respect to $C_4$, and by Observation \ref{pawleafedge}, we have:

\begin{observation}\label{obs:G3forcededges}
Every induced $G_3$ contains a forced edge of $G$, namely $x_4x_5$.
\end{observation}

{\sc Assumption 5.} By Observation \ref{obs:G3forcededges}, assume that the input graph is $G_3$-free: One can apply the
Edge Reduction to the corresponding forced edge of any induced $G_3$ (which can be done in polynomial time).

\medskip

Recall that by Observation \ref{dimC3C5C7C4} $(ii)$, every edge of an induced $C_4$ is excluded, and by Observation \ref{pawleafedge}, the leaf edge of an induced paw is excluded.

\begin{observation}\label{pawleafblackwhite}
Let $a,b,c,d$ induce a paw with $C_3$ $bcd$, and $ab \in E$. If $a$ is black then $b$ is white and $cd \in M$. That is, if the edge $cd$ is excluded, then $b$ is black and thus, vertex $a$ is forced to be white.
\end{observation}

{\sc Assumption 6.} By Observation \ref{pawleafblackwhite} and the fact that all $C_4$-edges are excluded, and the leaf edges of paws are excluded (by Observation \ref{pawleafedge}), assume that in the input graph, there is no induced paw with vertices $a,b,c,d$, $C_3$ $bcd$ and leaf edge $ab \in E$, such that $cd$ is a $C_4$-edge: One can apply the Vertex Reduction to vertex $a$ whenever $cd$ is a $C_4$-edge (which can be done in polynomial time).

\medskip

Let us conclude by pointing out that further similar observations could be stated to possibly detect forced edges or vertices forced to be white; for example, vertices forced to be white can be found by small degrees:

\begin{observation}\label{triangleneighbdeg1}
\mbox{ }
\begin{itemize}
\item[$(i)$] If $a,b,c$ induce a $C_3$ and $ad \in E$ such that $d_G(d)=1$ then $d$ is forced to be white.
\item[$(ii)$] If $a,b,c,d$ induce a $C_4$ and $d_G(d)=2$ then $d$ is forced to be white.
\end{itemize}
\end{observation}

{\bf Proof.}
$(i)$: Recall Observation \ref{pawleafedge}.

$(ii)$: Recall Observation \ref{dimC3C5C7C4} $(ii)$.
\qed

\medskip

Subsequently, concerning Assumptions 1-2, we will not recall them in an explicit way, but just recall that the input graph $G$ is $(K_4$, diamond, butterfly)-free. Concerning Assumptions 3-6, we will recall them in an explicit way.

\subsection{The distance levels of an $M$-edge $xy$ in a $P_3$}\label{subsec:distlevels}

Since it is trivial to check whether $G$ has a d.i.m.\ $M$ with exactly one edge, we can assume from now on that $|M| \geq 2$;
for an edge $xy \in E$, assume that $xy \in M$, and based on \cite{BraMos2017,BraMos2017/2}, we first describe some general structure properties for the distance levels of $xy \in M$. Since $G$ is $(K_4$, diamond, butterfly)-free, we have:

\begin{observation}\label{obse:neighborhood}
For every vertex $v$ of $G$, $N(v)$ is the disjoint union of isolated vertices and at most one edge. Moreover, for every edge $xy \in E$, there is at most one common neighbor of $x$ and $y$.
\end{observation}

We have:

\begin{observation}\label{obse:xy-in-P3}
If $|M| \geq 2$ then there is an edge in $M$ which is contained in an induced $P_3$ of $G$.
\end{observation}

{\bf Proof.}
Let $xy \in M$ and assume that $xy$ is not part of an induced $P_3$ of $G$. Since $G$ is connected and $|M| \geq 2$, $(N(x) \cup N(y)) \setminus \{x,y\} \neq \emptyset$, and since we assume that  $xy$ is not part of an induced $P_3$ of $G$ and $G$ is $K_4$- and diamond-free, there is exactly one neighbor of $xy$, namely a common neighbor, say $z$, of $x$ and $y$. Again, since $|M| \geq 2$, $z$ has a neighbor $a \notin \{x,y\}$, and since $G$ is $K_4$- and diamond-free, $a,x,y,z$ induce a paw. Clearly, the edge $za$ is excluded and has to be dominated by a second $M$-edge, say $ab \in M$ but now, since $G$ is butterfly-free, $zb \notin E$. Thus, $z,a,b$ induce a $P_3$ in $G$, and Observation \ref{obse:xy-in-P3} is shown.
\qed

\medskip

Thus, let $xy \in M$ be an $M$-edge for which there is a vertex $r$ such that $\{r,x,y\}$ induce a $P_3$ with edge $rx \in E$. Then, by the assumption that $xy \in M$, $x$ and $y$ are black and can lead to a (partial) feasible {\em $xy$-coloring} (if no contradiction arises).
We consider a partition of $V$ into the distance levels $N_i=N_i(xy)$, $i \ge 1$, (and $N_0:=\{x,y\}$) with respect to the edge $xy$ (under the assumption that $xy \in M$).

Recall that by (\ref{IV(M)partition}), $V=I \cup V(M)$ is a partition of $V$ where $I$ is an independent set (of white vertices) while $V(M)$ is the set of black vertices. Since we assume that $xy \in M$, clearly, $N_1 \subseteq I$ and thus:
\begin{equation}\label{N1subI}
N_1 \mbox{ is an independent set of white vertices.}
\end{equation}

Moreover, no edge between $N_1$ and $N_2$ is in $M$. Since $N_1 \subseteq I$ and all neighbors of vertices in $I$ are in $V(M)$, we have:
\begin{equation}\label{N2M2S2}
G[N_2] \mbox{ is the disjoint union of edges and isolated vertices.}
\end{equation}

Let $M_2$ denote the set of edges $uv \in E$ with $u,v \in N_2$ and let $S_2 = \{u_1,\ldots,u_k\}$ denote the set of isolated vertices in $N_2$; $N_2=V(M_2) \cup S_2$ is a partition of $N_2$. Obviously:
\begin{equation}\label{M2subM}
M_2 \subseteq M \mbox{ and } S_2 \subseteq V(M).
\end{equation}

If for $xy \in M$, an edge $e \in E$ is contained in {\bf every} dominating induced matching $M$ of $G$ with $xy \in M$, we say that $e$ is an {\em $xy$-forced} $M$-edge. The Edge Reduction step can also be applied for $xy$-forced $M$-edges (then, in the unsuccessful case, $G$ has no d.i.m.\ containing $xy$), and correspondingly for white vertices resulting from the black color of $x$ and $y$ and peripheral triangles.

\medskip

Obviously, by (\ref{M2subM}), we have:
\begin{equation}\label{M2xymandatory}
\mbox{Every edge in } M_2 \mbox{ is an $xy$-forced $M$-edge}.
\end{equation}

Thus, from now on, after applying the Edge Reduction for $M_2$-edges, we can assume that $M_2=\emptyset$, i.e., $N_2=S_2 = \{u_1,\ldots,u_k\}$. For every $i \in \{1,\ldots,k\}$, let $u'_i \in N_3$ denote the {\em $M$-mate} of $u_i$ (i.e., $u_iu'_i \in M$). Let $M_3=\{u_iu'_i: i \in \{1,\ldots,k\}\}$ denote the set of $M$-edges with one endpoint in $S_2$ (and the other endpoint in $N_3$). Obviously, by (\ref{M2subM}) and the distance condition for a d.i.m.\ $M$, the following holds:
\begin{equation}\label{noMedgesN3N4}
\mbox{ No edge with both ends in } N_3 \mbox{ and no edge between } N_3 \mbox{ and } N_4 \mbox{ is in } M.
\end{equation}

As a consequence of (\ref{noMedgesN3N4}) and the fact that every triangle contains exactly one $M$-edge (see Observation \ref{dimC3C5C7C4} $(i)$), we have:
\begin{equation}\label{triangleaN3bcN4}
\mbox{For every triangle $abc$} \mbox{ with } a \in N_3, \mbox{ and } b,c \in N_4, \mbox{ $bc \in M$ is an $xy$-forced $M$-edge}.
\end{equation}

This means that for the edge $bc$, the Edge Reduction can be applied, and from now on, we can assume that there is no such triangle $abc$ with $a \in N_3$ and $b,c \in N_4$, i.e., for every edge $uv \in E$ in $N_4$:
\begin{equation}\label{edgeN4N3neighb}
N(u) \cap N(v) \cap N_3 = \emptyset.
\end{equation}

\medskip

According to $(\ref{M2subM})$ and the assumption that $M_2=\emptyset$ (recall $N_2 = \{u_1,\ldots,u_k\}$), let:
\begin{enumerate}
\item[ ] $T_{one} := \{t \in N_3: |N(t) \cap N_2| = 1\}$;

\item[ ] $T_i := T_{one} \cap N(u_i)$, $i \in \{1,\ldots,k\}$;

\item[ ] $S_3 := N_3 \setminus T_{one}$.
\end{enumerate}

By definition, $T_i$ is the set of {\em private} neighbors of $u_i$ in $N_3$ (note that $u'_i \in T_i$),
$T_1 \cup \ldots \cup T_k$ is a partition of $T_{one}$, and $T_{one} \cup S_3$ is a partition of~$N_3$.

\begin{observation}\label{lemm:structure2}
The following statements hold:
\begin{enumerate}
\item[$(i)$] For all $i \in \{1,\ldots,k\}$, $T_i \cap V(M)=\{u_i'\}$.
\item[$(ii)$] For all $i \in \{1,\ldots,k\}$, $T_i$ is the disjoint union of isolated vertices and at most one edge.
\item[$(iii)$] $G[N_3]$ is bipartite.
\item[$(iv)$] $S_3 \subseteq I$, i.e., $S_3$ is an independent vertex set of white vertices.
\item[$(v)$] If a vertex $t_i \in T_i$ sees two vertices in $T_j$, $i \neq j$, $i,j \in \{1,\ldots,k\}$, then $u_it_i \in M$ is an $xy$-forced $M$-edge.
\end{enumerate}
\end{observation}

{\bf Proof.} $(i)$: Holds by definition of $T_i$ and by the distance condition of a d.i.m.\ $M$.

$(ii)$: Holds by Observation \ref{obse:neighborhood}.

$(iii)$: Follows by Observation \ref{dimC3C5C7C4} $(i)$ since every odd cycle in $G$ must contain at least one $M$-edge, and by (\ref{noMedgesN3N4}).

$(iv)$: If $v \in S_3$, i.e., $v$ sees at least two $M$-vertices, then clearly $v \in I$, and thus, $S_3 \subseteq I$ is an independent vertex set (recall that $I$ is an independent vertex set).

$(v)$: Suppose that $t_1 \in T_1$ sees $a$ and $b$ in $T_2$. Then, if $ab \in E$, $u_2,a,b,t_1$ induce a diamond in $G$. Thus, $ab \notin E$ and now,
$u_2,a,b,t_1$ induce a $C_4$ in $G$; the only possible $M$-edge for dominating $t_1a,t_1b$ is $u_1t_1$, i.e., $t_1=u'_1$.
\qed

\medskip

From now on, by Observation \ref{lemm:structure2} $(v)$, we can assume that for every $i,j \in \{1,\ldots,k\}$, $i \neq j$, any vertex $t_i \in T_i$ sees at most one vertex in $T_j$.

\medskip

See \cite{BraMos2017/2} for the following fact:

\begin{observation}\label{P1}
If $v \in N_2$ then $v$ is an endpoint of an induced $P_4$, say with vertices $v,v_1,v_2,v_3$, $v_1 \in N_1$, and with edges $vv_1 \in E$, $v_1v_2 \in E$, $v_2v_3 \in E$, and if $v \in N_i$ for $i \ge 3$ then $v$ is an endpoint of an induced $P_5$, say with vertices $v,v_1,v_2,v_3,v_4$ such that $v_1,v_2,v_3,v_4 \in \{x,y\} \cup N_1 \cup \ldots \cup N_{i-1}$ and with edges $vv_1 \in E$, $v_1v_2 \in E$, $v_2v_3 \in E$, $v_3v_4 \in E$.
\end{observation}

\noindent
{\bf Proof.}
First assume that $v \in N_2$. Since $xy$ is part of an induced $P_3$ with vertices $x,y,r$ and edges $xy, xr$, we have the following cases:
If $vr \in E$ then $(v,r,x,y)$ is a $P_4$. Now assume that $vr \notin E$, and let $v_1 \in N_1$ be a neighbor of $v$.
If $v_1r \in E$ then, since $G$ is diamond-free, $v_1x \notin E$ or $v_1y \notin E$. If $v_1x \in E$ and $v_1y \notin E$ then $(v,v_1,x,y)$ is a $P_4$, and
if $v_1x \notin E$ and $v_1y \in E$ then $(v,v_1,y,x)$ is a $P_4$.
Finally, if $v_1r \notin E$ then if $v_1x \in E$, $(v,v_1,x,r)$ is a $P_4$, and if $v_1x \notin E$ then $v_1y \in E$ and now, $(v,v_1,y,x)$ is a $P_4$.

If $v \in N_i$ for $i \ge 3$ then by similar arguments as above, $v$ is endpoint of a $P_5$ as described above.
\qed

\medskip

Let $X := \{x,y\} \cup N_1 \cup N_2 \cup N_3$ and $Y := V \setminus X$.
Subsequently, for checking if $G$ has a d.i.m.\ $M$ with $xy \in M$, we first consider the possible colorings for $G[X]$.

\section{Coloring $G[X]$}\label{ColoringG[X]}

As in \cite{BraMos2017/2}, we have:

\begin{proposition}\label{lemm:N4empty}
The following statements hold:
\begin{enumerate}
\item[$(i)$] For every edge $vw \in E$, $v,w \in N_3$, with $vu_i \in E$ and $wu_j \in E$ $($possibly $i=j)$, we have $|\{v,w\} \cap \{u'_i,u'_j\}| = 1$.
\item[$(ii)$] For every edge $st \in E$ with $s \in S_3$ and $t \in T_i$, $t=u'_i$ holds, and thus, $u_it$ is an $xy$-forced $M$-edge.
\end{enumerate}
\end{proposition}

{\bf Proof.} $(i)$: By (\ref{noMedgesN3N4}), $N_3$ does not contain any $M$-edge, and clearly, if $vw \in E$ then either $v$ or $w$ is black; without loss of generality, let $v$ be black but then $v=u'_i$ and $w$ is white, i.e., $w \neq u'_j$.

$(ii)$: By Observation \ref{lemm:structure2} $(iv)$, $S_3 \subseteq I$ and thus, by Proposition \ref{lemm:N4empty} $(i)$, for the edge $st$ with $s \in S_3$, $s$ is white and thus, $t=u'_i$ holds.
\qed

\medskip

From now on, after the Edge Reduction step, we can assume that each vertex in $S_3$ is isolated in $G[N_3]$. This means that every edge between $N_2$ and $N_3$ containing a vertex of $S_3$ is dominated. If $N_4 \neq \emptyset$ and $t \in N_4$ has a neighbor $s \in S_3$ then by (\ref{noMedgesN3N4}),  $t$ is forced to be black, and thus, every neighbor of $t$ in $N_3$ is forced to be white.

\medskip

Thus, for coloring $G[X]$, if $N_4 = \emptyset$ then we can assume that $S_3 = \emptyset$, i.e., $N_3=T_1 \cup \ldots \cup T_k$.

\medskip

Recall that:
\begin{itemize}
\item[$-$] All neighbors in $T_1 \cup \ldots \cup T_k$ of a black vertex in $T_1 \cup \ldots \cup T_k$ must be colored white, and all neighbors of a white vertex in $T_1 \cup \ldots \cup T_k$ must be colored black.

\item[$-$] Every $T_i$, $i \in \{1,\ldots,k\}$, contains exactly one vertex which is black. Thus, if $t_i \in T_i$ is black then all the remaining vertices of $T_i$ must be colored white.

\item[$-$] If all but one vertices of $T_i$, $i \in \{1,\ldots,k\}$, are white and the final vertex $t$ is not yet colored, then $t$ must be colored black. In particular, if  $|T_i|=1$, i.e., $T_i=\{t_i\}$, then $t_i$ is forced to be black, and $u_it_i$ is an $xy$-forced $M$-edge.

\item[$-$] All neighbors in $N_4$ of a black vertex in $T_1 \cup \ldots \cup T_k$ must be colored white, and all neighbors in $N_4$ of a white vertex in $T_1 \cup \ldots \cup T_p$ must be colored black.

\item[$-$] All neighbors in $T_1 \cup \ldots \cup T_k$ of a black vertex in $N_4$ must be colored white, and all neighbors in $T_1 \cup \ldots \cup T_p$ of a white vertex in $N_4$ must be colored black.

\item[$-$] If a white vertex in $T_i$ contacts $T_j$, $j \neq i$, then $T_j$ is completely colored.
\end{itemize}

\medskip

Recall that by Observation \ref{lemm:structure2} $(v)$ (and since by Assumption 2, $G$ is diamond-free), if $t_i \in T_i$ sees two vertices $t_j,t'_j \in T_j$, $i \neq j$, then $t_i$ is forced to be black and thus, all other vertices in $T_i$ are white, i.e., $T_i$ is completely colored. Moreover, $t_j,t'_j \in T_j$ are forced to be white. Thus, we can assume:
\begin{equation}\label{tiatmostoneneighbinIj}
\mbox{Every vertex of } T_i \mbox{ has at most one neighbor in } T_j \mbox{ for any } j \neq i.
\end{equation}

Let $K$ be a component of $G[S_2 \cup T_{one}]$ with $T_i = T_{one} \cap u_i$, $1 \le i \le p$, for the black vertex $u_i \in S_2$.

\medskip

Let us say that a vertex $t \in T_i$ (for $i \in \{1,\ldots,k\}$) is an {\em $N_3$-out-vertex} of $T_i$ if it contacts some $T_j$ with $j \neq i$,
$t$ is an {\em $N_4$-out-vertex} of $T_i$ if it contacts $N_4$, and $t$ is an {\em in-vertex} of $T_i$ otherwise.
For every $T_i$, the set of in-vertices of $T_i$ can be reduced to at most one such vertex:
\begin{equation}\label{atmostoneinvertex}
\mbox{ For every } i \in \{1,\ldots,k\}, T_i \mbox{ has at most one in-vertex}.
\end{equation}

In fact, if there is an in-vertex in $T_i$ then one can reduce the set of all in-vertices of $T_i$ to exactly one of them with minimum weight; that can be done in polynomial time.

\medskip

If $p=1$ then $K$ is {\em trivial}, and $K$ can be colored in polynomial time. Thus assume that $p \ge 2$.

\begin{proposition}\label{no3edgesbetweenTiTj}
If $xy \in M$ for a d.i.m.\ $M$ of $G$ then for any component $K$ in $G[S_2 \cup T_{one}]$, there are no three edges between $T_i$ and $T_j$, $i \neq j$.
\end{proposition}

{\em Proof.}
Suppose to the contrary that there are three edges between $T_1$ and $T_2$, say $t_1t_2 \in E$, $t'_1t'_2 \in E$, $t''_1t''_2 \in E$, $t_i,t'_i,t''_i \in T_i$, $i=1,2$. Then $t_1$ is black if and only if $t_2$ is white, $t'_1$ is black if and only if $t'_2$ is white, and $t''_1$ is black if and only if $t''_2$ is white. Without loss of generality, assume that $t_1$ is black, and $t_2$ is white. Then $t'_1$ is white, and $t'_2$ is black, but now, $t''_1$ and $t''_2$ are white which leads to a contradiction.

\medskip

Thus, Proposition \ref{no3edgesbetweenTiTj} is shown.
$\diamond$

\medskip

By Proposition \ref{no3edgesbetweenTiTj}, $xy \in M$ for a d.i.m.\ $M$ is impossible when for a component $K$ in $G[S_2 \cup T_{one}]$, there are three edges between $T_i$ and $T_j$, $i \neq j$. Thus assume that for each component $K$ in $G[S_2 \cup T_{one}]$, there are at most two edges between $T_i$ and $T_j$, $i \neq j$.

\begin{proposition}\label{N4neighborwith2inTi}
If $t \in N_4$ contacts two vertices $t_1,t_2 \in T_i$ then $t$ is black and $t_1,t_2$ are white.
\end{proposition}

{\em Proof.}
Recall that $T_i = T_{one} \cap N(u_i)$, $u_i$ is black, and $G$ is diamond-free, i.e., $t_1t_2 \notin E$. If $t$ is white then by (\ref{noMedgesN3N4}), $t_1$ and $t_2$ are black, i.e., $G$ has no d.i.m.\ $M$ with $xy \in M$. Thus, $t$ is black, by (\ref{noMedgesN3N4}), $t_1$ and $t_2$ are white, and
Proposition \ref{N4neighborwith2inTi} is shown.
$\diamond$

\begin{proposition}\label{S223frconncompinN3}
If there are two edges, namely between $T_i$ and $T_j$ and between $T_j$ and $T_{\ell}$ $($possibly $i=\ell)$ or between $T_i$ and $T_j$ and between $T_j$ and $N_4$ then they do not induce a $2P_2$ in $G[N_3 \cup N_4]$.
\end{proposition}

{\em Proof.}
Assume first without loss of generality that there are two edges, namely one between $T_1$ and $T_2$ and another between $T_2$ and $T_3$ or between $T_2$ and $N_4$.

Suppose to the contrary that $t_1t_2 \in E$ and $t'_2t_3 \in E$ induce a $2P_2$ in $G[N_3 \cup N_4]$ for $t_i \in T_i$, $i=1,2$, $t_3 \in T_3 \cup N_4$ and $t'_2 \in T_2$.
Recall that $y,x,r$ induce a $P_3$. If $u_2r \in E$ then $u_2,t_2,t_1,t'_2,t_3,r,x,y$ (with center $u_2$) induce an $S_{2,2,3}$, which is a contradiction. Thus, $u_2r \notin E$; let $w \in N_1$ with $u_2w \in E$ and without loss of generality, let $wx \in E$. But then $u_2,t_2,t_1,t'_2,t_3,w,x,r$ (with center $u_2$) induce an $S_{2,2,3}$, which is again a contradiction. Analogously, the same can be shown for two edges between $T_1$ and $T_2$.

\medskip

Thus, Proposition \ref{S223frconncompinN3} is shown.
$\diamond$

\medskip

Recall that $K$ is a component of $G[S_2 \cup T_{one}]$, and $t_i \in V(K) \cap T_{one}$ (without loss of generality, say $i = 1$, and $K$ is the subgraph induced by
$\{u_1,\ldots,u_p\}$ and by $T_1 \cup \ldots \cup T_p$, $2 \leq p \leq k$).

\begin{lemma}\label{DIMxyN1N2N3pol}
Showing that $G$ has no d.i.m.\ $M$ with $xy \in M$ and $t_1 \in V(M)$, or finding such a d.i.m.\ $M$ of $K$ with $u_1t_1 \in M$ can be done in polynomial time.
\end{lemma}

{\bf Proof.}
We will show that there is only a polynomial number of feasible black-white colorings of $K$.
The procedure starts with a subset $T_i$, say $i=1$. If $T_1$ is already completely colored, e.g.\ $|T_1|=1$ and $t_1 \in T_1$ is forced to be black, then
every neighbor $t_i$ of $t_1$ is white. If there is a forced white vertex $t \in T_i$ which contacts $T_1$ then its neighbor in $T_1$ is forced to be black, and if there are two forced white vertices $t \in T_i$, $t' \in T_j$,  which contact distinct vertices in $T_1$, say $tt_1 \in E$ and $t't'_1 \in E$, then $t_1$ and $t'_1$ are black, which is a contradiction by Observation \ref{lemm:structure2} $(i)$, i.e., $G$ has no d.i.m.\ $M$ with $xy \in M$.

Without loss of generality, we can assume that $T_1$ is not yet completely colored, and in particular, none of its vertices is already black.
Recall that by (\ref{tiatmostoneneighbinIj}), for every $j \in \{2,\ldots,p\}$, any vertex $t_1 \in T_1$ sees at most one vertex in $T_j$. We have to check for every vertex $t_1 \in T_1$ (which is not yet colored) whether $t_1$ could be black.

\medskip

Assume that $t_1 \in T_1$ is black. Then $T_1$ is completely colored, i.e., by Observation \ref{lemm:structure2} $(i)$, all vertices in $T_1 \setminus \{t_1\}$ are white. We are going to show that it will lead to a complete black-white coloring of all other vertices in $K$ or to a contradiction.

\medskip

First assume that the black vertex $t_1 \in T_1$ is an $N_3$-out-vertex, and recall that in any $T_j$, $t_1$ has at most one neighbor; without loss of generality, assume that $t_1$ has exactly one neighbor in $T_2$, say $t_1t_2 \in E$ with $t_2 \in T_2$.
If a white vertex $t'_1$ of $T_1$ has a neighbor $t'_2 \in T_2$ then $t'_2$ is black and all other vertices in $T_2$ are white, i.e., $T_2$ is completely colored. Thus assume that no white vertex in $T_1$ has a neighbor in $T_2$. Then for $t_1t_2 \in E$, $t_2$ is white, and if $t_2$ has a neighbor $t'_2 \in T_2$ then $t'_2$ is black and thus, $T_2$ is completely colored. Thus, assume that $t_2$ has no neighbor in $T_2$.
If all other vertices in $T_2$ are in-vertices then by $(\ref{atmostoneinvertex})$, there is only one of them, say $t'_2$, and thus, $t'_2$ is forced to be black, and $T_2$ is completely colored.

\medskip

Thus, first assume that there is a second $N_3$-out-vertex, say $t'_2 \in T_2$ which could be black but is not yet colored.
Then clearly, $t'_2$ has no neighbor in $T_1 \cup \{t_2\}$. Let $t'_2t_3 \in E$ for $t_3 \in T_3$. Again by (\ref{tiatmostoneneighbinIj}), $t_3t_2 \notin E$.
Since by Proposition \ref{S223frconncompinN3}, $t_1,t_2,t'_2,t_3$ do not induce a $2P_2$, the only possible edge between $t_1t_2$ and $t'_2t_3$ is $t_1t_3 \in E$,
but now, $t_3$ is white which implies that $t'_2$ is black and thus, $T_2$ is completely colored.

\medskip

Now assume that there is no other $N_3$-out-vertex in $T_2$. If $t'_2$ is an $N_4$-out-vertex, i.e., $t'_2t_3 \in E$ for $t_3 \in N_4$ then again, by Proposition \ref{S223frconncompinN3}, $t_1,t_2,t'_2,t_3$ do not induce a $2P_2$. Thus, $t_3t_1 \in E$ or $t_3t_2 \in E$.
If $t_3t_1 \in E$ then, by (\ref{noMedgesN3N4}), $t_3$ is white and now, $t'_2$ is black and thus, $T_2$ is again completely colored.
Thus assume that $t_3t_1 \notin E$ and $t_3t_2 \in E$. But then, by Proposition \ref{N4neighborwith2inTi}, $t_3$ is forced to be black, $t_2$ is forced to be white, and $t_1$ is forced to be black, which is a contradiction.

\medskip

Finally, if $t_1$ is no $N_3$-out-vertex in $T_1$ and we assume that $t_1$ is black then every $N_3$-out-vertex in $T_1$ is white, and thus,
every neighbor in $T_j$ of an $N_3$-out-vertex in $T_1$ is forced to be black, and thus, every $T_j$ which contacts $T_1$ is completely colored.

\medskip

In the same way as above, it can be done for every $T_i$ which is not yet completely colored but is adjacent to an already completely colored $T_j$.

\medskip

Thus, Lemma \ref{DIMxyN1N2N3pol} is shown.
\qed

\medskip

In particular, every component $K$ in $G[S_2 \cup T_{one}]$ with $K \cojoin N_4$ can be independently colored in polynomial time. Thus, if $N_4 = \emptyset$ then it is solved in polynomial time. From now on assume that $N_4 \neq \emptyset$.

\begin{remark}\label{rem1}
Recall that according to the proof of Lemma $\ref{DIMxyN1N2N3pol}$, once the color of a vertex in $T_i$, $1 \le i \le k$, is fixed to be black, then the color of all vertices of the component of $G[S_2 \cup T_{one}]$ containing $T_i$ is forced $($not necessarily in a feasible way$)$.
\end{remark}

By Remark \ref{rem1} and by Observation \ref{lemm:structure2}, we have:

\begin{proposition}\label{QconncompK}
Let $Q$ denote the family of components of $G[S_2 \cup T_{one}]$, and let $K$ be a member of $Q$.
\begin{itemize}
\item[$(i)$] By Observation $\ref{lemm:structure2}$~$(iv)$, any vertex in $V(K) \cap T_{one}$ contacting $S_3$ is black, and thus, the color of each vertex of $K$ is forced.

\item[$(ii)$] If for some $i \in \{1,\ldots,k\}$, $K$ contains a subset $T_i$ such that $|T_i| = 1$, then by Observation $\ref{lemm:structure2}$~$(i)$, the vertex in $T_i$ is black, and thus, the color of each vertex of $K$ is forced.

\item[$(iii)$] If for some $i \in \{1,\ldots,k\}$, $K$ contains a subset $T_i$ such that $|T_i| \geq 2$ and there is a vertex $z \in N_4$ with $z \join T_i$
 then, by the $C_4$-property in Observation $\ref{dimC3C5C7C4}$ $(ii)$ and since $G$ is diamond-free, $G$ has no d.i.m.\ with $xy \in M$.

\item[$(iv)$] If $K \cojoin N_4$ then clearly, $K$ can be colored independently of the other members of $Q$.
\end{itemize}
\end{proposition}

Thus, by Proposition \ref{QconncompK}, we can restrict $Q$ as follows: Let $Q^*$ be the family of components $K$ of $G[S_2 \cup T_{one}]$ such that:
\begin{itemize}
\item[(R1)] no vertex in $V(K) \cap T_{one}$ contacts $S_3$,
\item[(R2)] $V(K)$ contains no subset $T_i$, $1 \le i \le k$, such that $|T_i|= 1$,
\item[(R3)] for any $z \in N_4$, there is at least one non-neighbor of $z$ in $V(K) \cap N_3$, and
\item[(R4)] some vertex of $V(K)$ contacts $N_4$.
\end{itemize}

Next we show:

\begin{lemma}\label{lemm:G[X]coloring}
For $S_{2,2,3}$-free graphs $G$ with $N_4 \ne \emptyset$, the number of feasible $xy$-colorings of $G[X]$ $($with contact to $N_4)$ is at most polynomial.
In particular, such $xy$-colorings can be detected in polynomial time.
\end{lemma}

For the proof of Lemma \ref{lemm:G[X]coloring}, we will show the following proposition. 

\begin{proposition}\label{P2}
$|Q^*| \le 3$.
\end{proposition}

{\bf Proof.} Suppose to the contrary that $|Q^*| \ge 4$; let $L_1,\ldots,L_4 \in Q^*$.
Let $N_{x} := \{w \in N_1: w$ is adjacent to $x$ and non-adjacent to $y\}$,
$N_{y} := \{w \in N_1: w$ is adjacent to $y$ and non-adjacent to $x\}$, and $N_{xy} := \{w \in N_1: w$ is adjacent to $x$ and $y\}$.
Then $N_{xy} \cup N_x \cup N_y$ is a partition of $N_1$. Since $G$ is (diamond,$K_4$)-free, $|N_{xy}| \leq 1$. Since $\{r,x,y\}$ induce a $P_3$ with edge $rx \in E$,
we have $N_{x} \neq \emptyset$. We first show:

\begin{clai}\label{XQ*claim1}
For every $w \in N_x$ $(w \in N_y$, respectively$)$, at most one component in $Q^*$ contacts $w$.
\end{clai}

{\em Proof.}
Suppose to the contrary that there is a vertex $w \in N_x$ and there are two components $L_1,L_2 \in Q^*$ such that for $u_i \in S_2 \cap V(L_i)$, $i=1,2$, $wu_1 \in E$ and $wu_2 \in E$. Recall that $T_i=N(u_i) \cap T_{one}$ and $T_i \neq \emptyset$ (subsequently, we will also use this in the proofs of the following claims). Then by (R4), there are vertices $t_1 \in T_1$ and $z \in N_4$ with $t_1z \in E$, and by (R3), there is $t_2 \in T_2$ with $zt_2 \notin E$. But now, $w,x,y,u_2,t_2,u_1,t_1,z$ (with center $w$) induce an $S_{2,2,3}$, which is a contradiction. Thus, at most one component in $Q^*$ contacts $w \in N_x$. Analogously, it is true for $w \in N_y$.
$\diamond$

\begin{clai}\label{XQ*claim2}
At most two components in $Q^*$ contact $N_x$ $(N_y$, respectively$)$.
\end{clai}

{\em Proof.}
Suppose to the contrary that there are three such components $L_1,L_2,L_3 \in Q^*$ contacting $N_x$. By Claim \ref{XQ*claim1}, $L_1,L_2,L_3$ do not have common neighbors in $N_x$. Thus, let $w_i$, $i=1,2,3$, be distinct neighbors of $L_i$ in $N_x$, and let $u_i \in S_2 \cap V(L_i)$, $1 \le i \le 3$. But then $x,w_1,u_1,w_2,u_2,w_3,u_3,t_3$ (with center $x$) induce an $S_{2,2,3}$, which is a contradiction. Thus, at most two components in $Q^*$ contact $N_x$, and analogously, at most two components in $Q^*$ contact $N_y$.
$\diamond$

\medskip

However, we can show:

\begin{clai}\label{XQ*claim3}
If two components, say $L_1,L_2$, in $Q^*$ contact $N_x$ then no other component $L_i$, $i \ge 3$, in $Q^*$ contacts $N_y$.
Similarly, if two components, say $L_1,L_2$, in $Q^*$ contact $N_y$ then no other component $L_i$, $i \ge 3$, in $Q^*$ contacts $N_x$.
\end{clai}

{\em Proof.}
Assume that $L_1,L_2$ contact $N_x$, say $w_iu_i \in E$, $i=1,2$, for $w_i \in N_x$ and $u_i \in S_2 \cap V(L_i)$, $i=1,2$ (recall $w_1 \neq w_2$ and $w_1u_2 \notin E, w_2u_1 \notin E$ by Claim \ref{XQ*claim1}), and suppose to the contrary that there is a component $L_3 \in Q^*$ contacting $N_y$, say $wu_3 \in E$ for $w \in N_y$ and $u_3 \in S_2 \cap V(L_3)$.
By Claim \ref{XQ*claim1} and since $L_3$ contacts $w$, $L_1,L_2$ do not contact $w$ and $L_3$ does not contact $w_1,w_2$, i.e., $u_1w \notin E$, $u_2w \notin E$,
$u_3w_1 \notin E$, $u_3w_2 \notin E$, but now, $x,w_1,u_1,w_2,u_2,y,w,u_3$ (with center $x$) induce an $S_{2,2,3}$, which is a contradiction.
$\diamond$

\begin{clai}\label{XQ*claim4}
At most two components in $Q^*$ contact $N_{xy}$.
\end{clai}

{\em Proof.} Let $N_{xy}=\{w\}$ (recall that $|N_{xy}| \le 1$).
Suppose to the contrary that there are three such components $L_1,L_2,L_3 \in Q^*$ contacting $N_{xy}$, say $u_1,u_2,u_3$, $u_i \in S_2 \cap V(L_i)$, contact $w$.
Assume that $z \in N_4$ contacts $t_3 \in T_3$. Clearly, by (R3), there are vertices $t_1 \in T_1, t_2 \in T_2$ with $zt_1 \notin E$, and $zt_2 \notin E$.
But then $w,u_1,t_1,u_2,t_2,u_3,t_3,z$ (with center $w$) induce an $S_{2,2,3}$, which is a contradiction.
$\diamond$

\medskip

Next we show:
\begin{clai}\label{XQ*claim5}
If two components, say $L_1,L_2$, in $Q^*$ contact $N_{xy}$ then no other component $L_i$, $i \ge 3$, in $Q^*$ contacts $N_x$ or $N_y$.
\end{clai}

{\em Proof.} Let $N_{xy}=\{w\}$ (recall that $|N_{xy}| \le 1$), and let $L_1,L_2$ in $Q^*$ contact $w$, say $u_i \in S_2 \cap V(L_i)$, $i=1,2$, contact $w$.
Suppose to the contrary that there is an $L_3$ in $Q^*$ contacting $N_x$, say $u_3 \in S_2 \cap V(L_3)$ and $w_x \in N_x$ with $u_3w_x \in E$.
Recall $u_3w \notin E$ by Claim~\ref{XQ*claim4}.
If $u_1w_x \notin E$ and $u_2w_x \notin E$ then $w,u_1,t_1,u_2,t_2,x,w_x,u_3$ (with center $w$) would induce an $S_{2,2,3}$. By Claim \ref{XQ*claim1}, at most one of $u_1,u_2$ contacts $w_x$, say without loss of generality $u_1w_x \in E$ and $u_2w_x \notin E$. Recall that by (R3), there is a vertex $z \in N_4$ with $zt_1 \in E$ and $zt_2 \notin E$ but now, $u_1,t_1,z,w_x,u_3,w,u_2,t_2$ (with center $u_1$) induce an $S_{2,2,3}$, which is a contradiction.
$\diamond$

\medskip

From now on, we can assume that at most one component in $Q^*$ contacts $N_{xy}$. If no component in $Q^*$ contacts $N_{xy}$ then by Claims \ref{XQ*claim2} and
\ref{XQ*claim3}, $|Q^*| \le 3$. Thus assume that $L_1 \in Q^*$ contacts $N_{xy}$, say $u_1w \in E$ for $u_1 \in S_2 \cap V(L_1)$ and $N_{xy}=\{w\}$.

\medskip

Suppose to the contrary that there are three further components $L_2,L_3,L_4 \in Q^*$ contacting $N_x$, $N_y$. By Claim \ref{XQ*claim2}, at most two
of them contact $N_x$, and at most two of them contact $N_y$. Without loss of generality, assume that $L_2,L_3$ contact $N_x$, say $u_2w_2 \in E$ and $u_3w_3 \in E$ for $w_2,w_3 \in N_x$, $w_2 \neq w_3$ by Claim \ref{XQ*claim1}, and $u_i \in S_2 \cap V(L_i)$, $i=2,3$. By Claim~\ref{XQ*claim1}, $u_2w_3 \notin E$, $u_3w_2 \notin E$, and $u_1w_2 \notin E$, $u_1w_3 \notin E$. But then $x,w_2,u_2,w_3,u_3,w,u_1,t_1$ (with center $x$) induce an $S_{2,2,3}$, which is a contradiction.

\medskip

This finally leads to $|Q^*| \le 3$, and Proposition \ref{P2} is shown.
\qed

\medskip

The proof of Lemma \ref{lemm:G[X]coloring} follows by Remark 1 and by Proposition \ref{P2}.
\qed

\section{Coloring $G[Y]$}\label{ColoringG[Y]}

Recall that $X := \{x,y\} \cup N_1 \cup N_2 \cup N_3$ and $Y := V \setminus X$. From now on, let $Y \neq \emptyset$. Subsequently, we apply the polynomial-time solution for $S_{2,2,2}$-free graphs \cite{HerLozRieZamdeW2015} (see Theorem \ref{DIMpolresults} $(iii)$). In particular let us try to connect the ``coloring approach'' of \cite{HerLozRieZamdeW2015} with the above.
Recall that for a d.i.m.\ $M$ of $G$, $V(G)=V(M) \cup I$ is a partition of $V(G)$, all vertices of $V(M)$ are black and all vertices of $I$ are white.
Recall the following {\em forcing rules} (under the assumption that $xy \in M$):

\begin{itemize}
\item[$(i)$] If a vertex $v$ is white then all of its neighbors must be black.
\item[$(ii)$] If two adjacent vertices are black then all of their neighbors are white.
\item[$(iii)$] If a vertex $u$ is black and all of its neighbors, except $v \in N(u)$, are white, then $v$ must be black.
\end{itemize}

Let us fix a feasible $xy$-coloring of $X$ if there is one (otherwise $xy \notin M$). Consequently, $G[Y]$ has a fixed partial $xy$-coloring of its vertices, due to the forcing rules. We try to extend it to a complete feasible $xy$-coloring.

\begin{proposition}\label{obse:N4}
The fixed $xy$-coloring of $G[X]$ leads to a unique coloring of all vertices of $N_4$.
\end{proposition}

{\bf Proof.} Let $v \in N_4$ and let $u \in N_3$ be a neighbor of $v$. If $u$ is white then $v$ is black, and if $u$ is black then, since
by fact (\ref{noMedgesN3N4}), the $M$-mate of $u$ is in $N_2$, $v$ is white.
\qed

\medskip

First assume that $G[Y]$ is $S_{2,2,2}$-free. Then by Theorem \ref{DIMpolresults} $(iii)$ (and the coloring approach of \cite{HerLozRieZamdeW2015}), one can check in polynomial time whether $G[Y]$ admits a feasible coloring of its vertices (consistent with the fixed $xy$-coloring of $X$), i.e., whether $G$ admits a complete $xy$-coloring of its vertices.

\medskip

From now on, assume that $G[Y]$ is not $S_{2,2,2}$-free. Then let us show that, while $G[Y]$ contains an induced $S_{2,2,2}$, say $H$, one can remove some vertices of $H$, in order to obtain a reduced subgraph of $G[Y]$ which admits a feasible coloring of its vertices (consistent with the fixed $xy$-coloring of $X$) if and only if $G[Y]$ does so.

\medskip

Let $H$ be an induced $S_{2,2,2}$ in $G[Y]$ with vertices $V(H)=\{d,a_1,a_2,b_1,b_2,c_1,c_2\}$ and edges $da_1$, $db_1$, $dc_1$, $a_1a_2$, $b_1b_2$, $c_1c_2$
(in particular, $d$ is the center of $H$).
Let $p := \min \{i: N_i \cap V(H) \neq \emptyset\}$, that is, $N_p$ is the $xy$-distance level with smallest distance to $xy$ to which a vertex of $H$ belongs (in particular $p \geq 4$ by construction).

\medskip

Then there exists a vertex, say $z \in N_{p-1}$, $p-1 \ge 3$, contacting $H$. By Observation \ref{P1}, we have:

\begin{itemize}
\item[$-$] $z$ is the endpoint of an induced $P_5$ $(z,z_2,z_3,z_4,z_5)$, say $P(z)$, such that $z_2 \in N_{p-2}, z_3 \in N_{p-3}, z_4 \in N_{p-4}$\\
(note that then no vertex in $H$ is adjacent to $z_2,z_3,z_4,z_5$, and no neighbor of $H$ is adjacent to $z_3,z_4,z_5$).
\end{itemize}

\begin{proposition}\label{P3}
Vertex $z$ is nonadjacent to $d$, and in general, $N_{p-1} \cap N(d) = \emptyset$.
\end{proposition}

{\bf Proof.} Suppose to the contrary that $z$ is adjacent to $d$. If $z \cojoin \{a_1,a_2\}$ and $z \cojoin \{b_1,b_2\}$ then $d,a_1,a_2,b_1,b_2,z,z_2,z_3$ (with center $d$) induce an $S_{2,2,3}$, and analogously for $z \cojoin \{a_1,a_2\}$ and $z \cojoin \{c_1,c_2\}$, as well as for $z \cojoin \{b_1,b_2\}$ and $z \cojoin \{c_1,c_2\}$. Thus, without loss of generality, we can assume that $z$ sees $a_1$ or $a_2$ and $z$ sees $b_1$ or $b_2$.
Now, since $G$ is diamond-free, $z$ is adjacent to exactly one vertex in $\{a_1,a_2\}$ and to exactly one vertex in $\{b_1,b_2\}$,
but then $z,a_1,a_2,b_1,b_2,z_2,z_3,z_4$ (with center $z$) induce an $S_{2,2,3}$, which is a contradiction.
\qed

\begin{proposition}\label{P4}
Vertex $z$ is adjacent to some vertex in $\{a_1,b_1,c_1\}$.
\end{proposition}

{\bf Proof.} Suppose to the contrary that $z \cojoin \{a_1,b_1,c_1\}$. Then, since by Proposition \ref{P3}, $zd \notin E$,
$z$ is adjacent to some vertex in $\{a_2,b_2,c_2\}$.

If $z$ is adjacent to exactly one vertex in $\{a_2,b_2,c_2\}$, say by symmetry, $za_2 \in E$, $zb_2 \notin E$, $zc_2 \notin E$, then
$d,a_1,a_2,z,b_1,b_2,c_1,c_2$ (with center $d$) induce an $S_{2,2,3}$.

If $z$ is adjacent to at least two vertices in $\{a_2,b_2,c_2\}$, say by symmetry, $za_2 \in E$, $zb_2 \in E$, then $z,a_1,a_2,b_1,b_2,z_2,z_3,z_4$ (with center $z$) induce an $S_{2,2,3}$, which is a contradiction.
\qed

\begin{proposition}\label{P5}
Without loss of generality, we can assume that $z \join \{a_1,a_2\}$, $z$ sees exactly one vertex in $\{b_1,b_2\}$, and $z \cojoin \{c_1,c_2\}$.
\end{proposition}

{\bf Proof.} By Proposition \ref{P4}, assume without loss of generality that $za_1 \in E$.
Since $d,b_1,b_2,c_1,c_2,a_1,z,z_2$ (with center $d$) do not induce an $S_{2,2,3}$, $z$ is adjacent either to some vertex in $\{b_1,b_2\}$ or to some vertex in $\{c_1,c_2\}$; without loss of generality, let $z$ be adjacent to some vertex in $\{b_1,b_2\}$. Then, since $z,a_1,a_2,b_1,b_2,z_2,z_3,z_4$ (with center $z$) do not induce an $S_{2,2,3}$, either $z \join \{a_1,a_2\}$ or $z \join \{b_1,b_2\}$;
without loss of generality, let $z \join \{a_1,a_2\}$.

Now, since $G$ is butterfly-free, neither $z \join \{b_1,b_2\}$ nor $z \join \{c_1,c_2\}$, and again, by the previous arguments, $z$ has a neighbor in $\{b_1,b_2\}$ or in $\{c_1,c_2\}$; by symmetry, let $z$ have exactly one neighbor in $\{b_1,b_2\}$, say $b_i$ where $i \in \{1,2\}$. Moreover, if $z$ has a neighbor in $\{c_1,c_2\}$, say $c_j$ where $j \in \{1,2\}$, then $z,b_1,b_2,c_1,c_2,z_2,z_3,z_4$ (with center $z$) induce an $S_{2,2,3}$. Thus, $z \cojoin \{c_1,c_2\}$, and Proposition \ref{P5} is shown.
\qed

\medskip

By Observation \ref{dimC3C5C7C4} $(i)$, it follows:

\begin{proposition}\label{P5one}
Exactly one of the edges $za_1,za_2,a_1a_2$ of the triangle $za_1a_2$ is in $M$.
\end{proposition}

Now let us distinguish between a facilitated case, namely in which $N_7$ is empty, and the general case, namely in which $N_7$
may be non-empty, where the facilitated case will be used as a sub-procedure.

\section{The facilitated case in which $N_7$ is empty}

\begin{lemma}\label{lemm:N7}
If $N_7 = \emptyset$ then one can detect a white vertex of $H$ in polynomial time.
\end{lemma}

{\bf Proof.} Assume $N_7 = \emptyset$. Then $p \le 6$, hence $z \in N_j$ with $3 \le j \le 5$.
If $z \in N_5$ then $V(H) \subseteq N_6$ since $N_7 = \emptyset$, i.e., $d \in N_6$ but by Proposition \ref{P3}, $N_{p-1} \cap N(d) = \emptyset$, which is a contradiction. Thus, $z \in N_3 \cup N_4$. Then by the fixed $xy$-coloring of $G[X]$ (if $z \in N_3$) and by Proposition \ref{obse:N4} (if $z \in N_4$), the color of $z$ is fixed. Now, since $z,a_1,a_2$ induce a $C_3$, we have:

\begin{itemize}
\item[$-$] if $z$ is black and $zb_2 \in E$, then $b_2$ is white;
\item[$-$] if $z$ is black and $zb_1 \in E$, then $b_1$ is white;
\item[$-$] if $z$ is white, then $d$ is white.
\end{itemize}

Finally, vertex $z$ can be computed in polynomial time by definition.
\qed

\begin{lemma}\label{theo:N7}
If $G$ is $S_{2,2,3}$-free and for $xy \in E$ which is part of an induced $P_3$ in $G$, $N_7 = \emptyset$, then one can check in polynomial time if $G$ has a d.i.m.\ $M$ with $xy \in M$.
\end{lemma}

{\bf Proof.} The proof is given by the following procedure.

\begin{proc}[DIM-with-$xy$-$N_1$-$N_6$]\label{DIMwithxyN1N6}

\begin{tabbing}	
xxxxxxx \= \kill\\
{\bf Input:} \> A connected $(S_{2,2,3},K_4$,diamond,butterfly$)$-free graph $G = (V,E)$, which satisfies\\
\> Assumptions 1-6 of Section $\ref{exclforced}$ and\\
\> an edge $xy \in E$, which is part of an induced $P_3$ in $G$, with $N_7(xy)=\emptyset$.\\
{\bf Task:} \> Return a d.i.m.\ $M$ with $xy \in M$ $($STOP with success$)$ or\\
\> a proof that $G$ has no d.i.m.\ $M$ with $xy \in M$ $($STOP with failure$)$.
\end{tabbing}

\begin{itemize}

\item[$(a)$] Set $M:= \{xy\}$. Determine the distance levels $N_i = N_i(xy)$, $1 \le i \le 6$, with respect to $xy$.

\item[$(b)$] Check whether $N_1$ is an independent set $($see fact $(\ref{N1subI}))$ and $N_2$ is the disjoint union of edges and isolated vertices $($see fact $(\ref{N2M2S2}))$. If not, then STOP with failure.

\item[$(c)$]  For the set $M_2$ of edges in $N_2$, apply the Edge Reduction for every edge in $M_2$. Moreover, apply the Edge Reduction for each edge $bc$ according to fact $(\ref{triangleaN3bcN4})$ and then for each edge $u_it_i$ according to Observation $\ref{lemm:structure2}$ $(v)$.

\item[$(d)$]  {\bf if} $N_4 = \emptyset$ {\bf then} apply the approach described in Section $\ref{ColoringG[X]}$. Then either return that $G$ has no d.i.m.\ $M$ with $xy \in M$ or return $M$ as a d.i.m.\ with $xy \in M$.

\item[$(e)$]  {\bf if} $N_4 \neq \emptyset$ {\bf then} for $X := \{x,y\} \cup N_1 \cup N_2 \cup N_3$ and $Y := V \setminus X$ $($according to the results of Section  $\ref{ColoringG[X]})$ {\bf do}

\begin{itemize}
\item[$(e.1)$] Compute all black-white $xy$-colorings of $G[X]$. If no such $xy$-coloring without contradiction exists, then STOP with failure.

\item[$(e.2)$]  {\bf for each} $xy$-coloring of $G[X]$  {\bf do}
\begin{enumerate}
\item[$(e.2.1)$]  Derive a partial coloring of $G[Y]$ by the forcing rules;
\item[ ] {\bf if} a contradiction arises in vertex coloring {\bf then} STOP with failure for this $xy$-coloring of $G[X]$ and proceed to the next $xy$-coloring of $G[X]$.

\item[$(e.2.2)$] Set $G[Y]: = F$.

\item[$(e.2.3)$] {\bf while} $F$ contains a $S_{2,2,2}$ say $H$ {\bf do}:
\begin{itemize}
\item[$(i)$] Detect a white vertex $h \in V(H)$ by Lemma $\ref{lemm:N7}$, and
\item[$(ii)$] apply the Vertex Reduction to $h$;
\item[ ] {\bf if} a contradiction arises in the vertex coloring {\bf then} STOP with failure for this $xy$-coloring of $G[X]$ and proceed to the next $xy$-coloring of $G[X]$ {\bf else} let $F'$ be the resulting subgraph of $F$; set $F:= F'$.
\end{itemize}
\item[$(e.2.4)$] Apply the algorithm of Hertz et al.\ $($see Theorem $\ref{DIMpolresults}$ $(iii))$ to determine if $F$ $\{$which is $S_{2,2,2}$-free by the above$\}$ has a d.i.m.
\item[ ] {\bf if} $F$ has a d.i.m.\ {\bf then} STOP and return the $xy$-coloring of $G$ derived by the $xy$-coloring of $G[X]$ and by such a d.i.m.\ of $G[Y]$.
\end{enumerate}

\item[$(e.3)$] STOP and return ``$G[Y]$ has no d.i.m.''.
\end{itemize}
\end{itemize}
\end{proc}

The correctness of Procedure \ref{DIMwithxyN1N6} follows from the structural analysis of $S_{2,2,3}$-free graphs with a d.i.m.\ and by the results in the present section.

\medskip

The polynomial time bound of Procedure \ref{DIMwithxyN1N6} follows from the fact that Steps (a) and (b) can clearly be done in polynomial time, Step (c) can be done in polynomial time since the Edge Reduction can be done in polynomial time, Step (d) can be done in polynomial time by the results in Section $\ref{ColoringG[X]}$, Step (e) can be done in polynomial time since the Vertex Reductions can be executed in polynomial time, since the solution algorithm of Hertz et al. (see Theorem \ref{DIMpolresults} $(iii)$) can be executed in polynomial time, and by the results in the present section.
\qed

\medskip

Now we consider the general case when $N_7$ may be nonempty. 


\section{The general case in which $N_7$ may be non-empty}\label{sec:N7nonempty}

By Proposition \ref{P5}, let us distinguish between $zb_2 \in E$ and $zb_1 \in E$.
Then the goal is to detect a white vertex of $S_{2,2,2}$ $H$ or a peripheral triangle with a vertex of $H$. Recall $p := \min \{i: N_i \cap V(H) \neq \emptyset\}$.

\begin{proposition}\label{obse:p5}
If $p \leq 5$, then one can easily detect a white vertex of $H$.
\end{proposition}

{\bf Proof.} In fact, if $p \leq 5$, then the color of $z$ (recall $z \in N_{p-1}$) is known by Proposition \ref{obse:N4}. Then, as in the proof of Lemma \ref{lemm:N7}, one can easily detect a white vertex of $H$.
\qed

\medskip

From now on, let us assume that $p \geq 6$. In particular, let $z_5 \in N_{p-5}$ be a neighbor of $z_4$, and let $z_6 \in N_{p-6}$ be a neighbor of $z_5$.

\subsection{The case $zb_2 \in E$}
If $b_2$ has two neighbors $m_1,m_2 \notin V(H) \cup \{z\}$ such that $b_2,m_1,m_2$ induce a triangle then we call $b_2,m_1,m_2$ an {\em external triangle}.

\begin{proposition}\label{b2triangle}
If $b_2$ is part of an external triangle $b_2,m_1,m_2$ then $b_2$ is white and $m_1m_2 \in M$ is forced.
\end{proposition}

{\bf Proof.} Let $b_2,m_1,m_2$ be an external triangle. Recall that $z,b_2,b_1,d,a_1$ induce a $C_5$.
By Observation \ref{pawleafedge}, the (leaf) edges $zb_2$, $a_1d$, and $b_2b_1$ are excluded. Then by Observation \ref{dimC3C5C7C4} $(i)$, for the $C_5$, either $b_1d \in M$ or $a_1z \in M$ which implies that $b_2$ is white and consequently, $m_1m_2 \in M$ is forced.
Thus, Proposition \ref{b2triangle} is shown.
\qed

\begin{proposition}\label{degb2>2dblackimpl}
If $d_G(b_2)>2$ and $b_2$ is not part of an external triangle then $d$ is white.
\end{proposition}

{\bf Proof.} Let $m_1 \notin \{z,b_1\}$ be a third neighbor of $b_2$. Suppose to the contrary that $d$ is black. Then, since $a_1d$ is excluded, $a_1$ is white which implies that $za_2 \in M$, $z_2$ is white, $z_3$ is black, $b_2$ is white and thus, $m_1$ and $b_1$ are black, i.e., $b_1d \in M$.
Moreover, $c_1$ is white and $c_2$ is black. Clearly, $m_1$ misses $b_1,d,z,a_2$, and recall that $m_1$, as a neighbor of $H$, misses $z_3,z_4,z_5$

\medskip

Let $m_2$ be an $M$-mate of $m_1$. Since by assumption, $b_2$ is not part of an external triangle, we have $m_2b_2 \notin E$. We first claim that
$m_1c_2 \notin E$, or equivalently, $m_2 \neq c_2$:

If $m_2 = c_2$ then, since $b_2,m_1,c_2,b_1,d,z,z_2,z_3$ (with center $b_2$) do not induce an $S_{2,2,3}$, we have $m_1z_2 \in E$ but now,
$z_2,m_1,c_2,z,a_2,z_3,z_4,z_5$ (with center $z_2$) induce an $S_{2,2,3}$, which is a contradiction. Thus, $m_2 \neq c_2$ is shown.

\medskip

Recall again that $m_1$, as a neighbor of $H$, misses $z_3,z_4,z_5$; similarly, $m_2$ misses $z_4,z_5$; furthermore, $m_2$ misses $z_3$ as well, since $z_3$ is black.

\medskip

Since $V(H) \cup \{m_1\}$ does not induce an $S_{2,2,3}$, we have $m_1a_1 \in E$ or $m_1c_1 \in E$.

\medskip

Next we claim that $m_1z_2 \notin E$:

\medskip

Suppose to the contrary that $m_1z_2 \in E$. If $m_1c_1 \in E$ then $m_1,c_1,c_2,b_2,b_1,z_2,z_3,z_4$ (with center $m_1$) would induce an $S_{2,2,3}$, and if $m_1a_1 \in E$ then $m_1,a_1,a_2,b_2,b_1,z_2,z_3,z_4$ (with center $m_1$) would induce an $S_{2,2,3}$, which is a contradiction. Thus $m_1z_2 \notin E$.

Since $b_2,m_1,m_2,b_1,d,z,z_2,z_3$ (with center $b_2$) do not induce an $S_{2,2,3}$, we have $z_2m_2 \in E$ (recall $z_3m_2 \not \in E$).

But then $z_2,m_2,m_1,z,a_2,z_3,z_4,z_5$ (with center $z_2$) induce an $S_{2,2,3}$, which is a contradiction.

\medskip

Thus, Proposition \ref{degb2>2dblackimpl} is shown.
\qed

\begin{proposition}\label{b2deg2dwhite}
If $d_G(b_2)=2$ then $d$ is white or there is a peripheral triangle with $c_1$ and $c_2$.
\end{proposition}

{\bf Proof.}
Let $d_G(b_2)=2$, i.e., $N_G(b_2)=\{b_1,z\}$. Note that in this case, $b_1$ must be black since $zb_2 \notin M$ and the edge $b_1b_2$ can only be dominated by an $M$-edge containing vertex $b_1$.

\medskip

Suppose to the contrary that $d$ is black, i.e., $b_1d \in M$, and there is no peripheral triangle with $c_1$ and $c_2$.

\medskip

Then $a_1$ is white and thus, $za_2 \in M$ which implies that also $b_2,c_1,z_2$ are white and $c_2,z_3$ are black, and then there is an $M$-mate $c \notin V(H)$ such that $cc_2 \in M$.
Clearly, since $d_G(b_2)=2$, $cb_2 \notin E$, and since $c$ is black, $c \cojoin \{a_2,z,b_1,d\}$.
Moreover, since $c$ contacts $H$, $c$ misses $z_3,z_4,z_5$, and $cz_2 \notin E$ since $z_2,z,a_2,c,c_2,z_3,z_4,z_5$ (with center $z_2$)
do not induce an $S_{2,2,3}$.

Since $V(H) \cup \{c\}$ does not induce an $S_{2,2,3}$ with center $d$, we have $ca_1 \in E$ or $cc_1 \in E$.
Since $a_1,c,c_2,d,b_1,z,z_2,z_3$ (with center $a_1$) do not induce an $S_{2,2,3}$, we have $ca_1 \notin E$, which implies $cc_1 \in E$.

Next we claim:
\begin{equation}\label{d(c1)=3}
N(c_1)=\{d,c_2,c\} \mbox{, i.e., } d_G(c_1)=3.
\end{equation}

{\em Proof.}
Suppose to the contrary that there is a vertex $d_1 \notin V(H) \cup \{c\}$ with $c_1d_1 \in E$. Then since $c_1$ is white, $d_1$ is black and thus, there is an $M$-mate $d_2$ of $d_1$. Since $za_2, db_1, cc_2 \in M$, we have $d_2 \notin \{z,a_2,d,b_1,c,c_2\}$.

Since $c_1,c,c_2,d_1,d_2$ do not induce a butterfly (recall that $G$ is butterfly-free), $d_2c_1 \notin E$.
Clearly, since $d,a_1,a_2,b_1,b_2,c_1,d_1,d_2$ (with center $d$) do not induce an $S_{2,2,3}$, and since $d_G(b_2)=2$ and $b_1,d,a_2$ are black, we have $d_1a_1 \in E$ or $d_2a_1 \in E$.

Since $z,a_2,a_1,d_1,d_2$ do not induce a butterfly, we have $d_1a_1 \notin E$ or $d_2a_1 \notin E$.
If $d_2a_1 \in E$ and $d_1a_1 \not \in E$, then $d,c_1,d_1,d_2,a_1$ and $z,a_2,c_2,c$ induce a $G_1$, which is not possible by Assumption 3. Thus $d_2a_1 \notin E$ and $d_1a_1 \in E$.

Recall that $z$ is black, $z_2$ is white and $z_3$ is black, and $d_1$ (as a neighbor of $H$) misses $z_3,z_4,z_5$.
Since $a_1,d_1,d_2,d,b_1,z,z_2,z_3$ (with center $a_1$) do not induce an $S_{2,2,3}$, we have $d_1z_2 \in E$ or $d_2z_2 \in E$.
If $d_1z_2 \in E$ then $d_1,c_1,c_2,a_1,a_2,z_2,z_3,z_4$ (with center $d_1$) would induce an $S_{2,2,3}$. Thus, $d_1z_2 \notin E$ which implies $d_2z_2 \in E$ but now, $d_1,c_1,c_2,a_1,a_2,d_2,z_2,z_3$ (with center $d_1$) induce an $S_{2,2,3}$, which is a contradiction. Thus, (\ref{d(c1)=3}) is shown.
$\diamond$

\medskip

Since we supposed that there is no peripheral triangle with $c_1$ and $c_2$, by (\ref{d(c1)=3}), we have a further neighbor of $c$ or $c_2$, without loss of generality, say $q \notin V(H) \cup \{c\}$ with $cq \in E$.

Clearly, $q$ is white. Thus, $q$ misses $a_1,z_2,b_2,c_1$, and since $G$ is diamond-free, $q$ misses $c_2$.

Since $d,a_1,a_2,b_1,b_2,c_1,c,q$ (with center $d$) do not induce an $S_{2,2,3}$, vertex $q$ is adjacent to $b_1$ or $d$ or $a_2$. It follows that $q$ contacts $H$.
Thus, $q$ misses $z_3,z_4,z_5$ by definition of $z$.

If $qa_2 \in E$ then $qb_1 \in E$ or $qd \in E$ since $d,b_1,b_2,c_1,c_2,a_1,a_2,q$ (with center $d$) do not induce an $S_{2,2,3}$.

If $qb_1 \in E$ and $qd \notin E$ then $qz \in E$ since $d,b_1,q,c_1,c_2,a_1,z,z_2$ (with center $d$) do not induce an $S_{2,2,3}$.
But now, $z,z_2,z_3,a_1,d,q,c,c_2$ (with center $z$) induce an $S_{2,2,3}$, which is a contradiction.

If $qb_1 \notin E$ and $qd \in E$ then $qa_2 \in E$ since $d,b_1,b_2,a_1,a_2,q,c,c_2$ (with center $d$) do not induce an $S_{2,2,3}$.
If $qa_2 \in E$ then, since $G$ is diamond-free, $qz \notin E$.
But now, $q,c,c_2,d,b_1,a_2,z,z_2$ (with center $q$) induce an $S_{2,2,3}$, which is a contradiction.

Thus, we have $qb_1 \in E$ and $qd \in E$, which leads to a $C_4$ induced by $q,c,c_1,d$ and a triangle induced by $q,d,b_1$, and then to a paw
induced by $q,d,b_1,b_2$ with the $C_4$-edge $qd$ but this is not possible by Assumption 6.

Thus, the assumption that $d$ is black and there is no peripheral triangle with $c_1$ and $c_2$ leads to a contradiction, and
Proposition \ref{b2deg2dwhite} is shown.
\qed

\begin{lemma}\label{lemm:zb2}
If $zb_2 \in E$, then we can detect a white vertex of $H$ or a peripheral triangle with a vertex in $H$ in polynomial time.
\end{lemma}

{\bf Proof.}
Checking if there is a peripheral triangle with a vertex in $H$ can be done in polynomial time. Then let us assume that there is no such peripheral triangle---in particular, there is no peripheral triangle with $c_1$ and $c_2$.

\medskip

If $b_2$ is part of an external triangle, then by Proposition \ref{b2triangle}, $b_2$ is white. If $b_2$ is not part of an external triangle, then we have:
If $d_G(b_2)>2$, then by Proposition \ref{degb2>2dblackimpl}, $d$ is white; if $d_G(b_2)=2$ then by Proposition \ref{degb2>2dblackimpl} and since there is no peripheral triangle with $c_1$ and $c_2$, $d$ is white. Obviously, we can check all of these steps in polynomial time.
\qed

\subsection{The case $zb_1 \in E$}\label{section5.2}

Recall that by Proposition \ref{obse:p5}, we can assume that for $z \in N_{p-1}$, $p \geq 6$.
Since $zb_1 \in E$, vertices $z,a_1,d,b_1$ induce a $C_4$, by Observation \ref{dimC3C5C7C4} $(ii)$, $za_1$ is excluded, and by Observation \ref{dimC3C5C7C4} $(i)$, either $za_2 \in M$ or $a_1a_2 \in M$, i.e., $a_2$ is black.

\medskip

If $a_2$ has a third neighbor, say $x \notin \{z,a_1\}$ then, by Observation \ref{pawleafblackwhite}, the paw induced by $z,a_1,a_2,x$
would imply that vertex $x$ is white (which leads to Vertex Reduction). Thus, from now on, we can assume that $d_G(a_2)=2$.

\medskip

Let $P(H) := \{t \in V \setminus V(H): t$ contacts $H$, and $t$ is the endpoint of an induced $P_4$ of $G[V \setminus V(H)]$ of vertices $t,t_2,t_3,t_4$ such that $t_2,t_3,t_4$ do not contact $H \}$.

\medskip

Clearly, $z \in P(H)$.

\begin{proposition}\label{prop:adjacencyH}
Let $t \in P(H)$ and let $W := \{(a_1,a_2),(b_1,b_2),(c_1,c_2)\}$. Then $t$ is nonadjacent to $d$, is adjacent to both vertices of one pair in $W$, is adjacent to exactly one vertex of one pair in $W$, and is nonadjacent to both vertices of one pair in $W$.
\end{proposition}

{\bf Proof.} Proposition \ref{prop:adjacencyH} follows by Propositions \ref{P3}, \ref{P4}, and \ref{P5}, since they hold for $z$ and (by a similar argument) for any vertex of $P(H)$ as well.
\qed

\medskip

Since $G$ is (diamond,$K_4$)-free, Proposition \ref{prop:adjacencyH} implies $|P(H)| \le 3$.
Note that $P(H)$ can be computed in polynomial time.

\begin{lemma}\label{prop:P(H)second}
If $P(H) \setminus \{z\} = \emptyset$ (i.e. if $P(H) = \{z\}$) then:
\begin{itemize}
  \item [$(i)$] $z$ is a cut-vertex for $G$; in particular let $K$ denote the component of $G[V \setminus \{z\}]$ containing $H$, and let $K_{xy}$ denote the component of $G[V \setminus \{z\}]$ containing $xy$;
  \item [$(ii)$] in $G[V(K) \cup \{z\}]$ we have $dist(za_2,v) \leq 6$ and $dist(a_1a_2,v) \leq 6$ for any $v \in K$;
  \item [$(iii)$] one can check in polynomial time $-$ by Lemma \ref{theo:N7} $-$ whether $G[V(K) \cup \{z\}]$ admits a d.i.m.\ containing $za_2$ and a d.i.m.\ containing $a_1a_2$.
  \item [$(iv)$] every component of $G[V \setminus \{z\}]$ except $K$ and $K_{xy}$ contains only one vertex, and if there is such a component, say $\{z'\}$, then $z'$ is white.
\end{itemize}
\end{lemma}

{\bf Proof.} $(i)$: Recall $X := \{x,y\} \cup N_1 \cup N_2 \cup N_3$. By definition of $P(H)$, since $P(H) = \{z\}$ and since $p \geq 6$, each path from $H$ to $X$ must involve $z$. Then $G[V \setminus \{z\}]$ is disconnected: in particular, $G[X]$ and $H$ belong to two distinct components of it.

\medskip

$(ii)$: We first claim that in $G[V(K) \cup \{z\}]$ we have $dist(za_2,v) \leq 6$ for any $v \in K$. Suppose to the contrary that there is a vertex $v \in K$ such that $dist(za_2,v) \geq 7$. Then clearly $v \not \in H$. Let $P$ be a shortest path in $K$ from $v$ to $H$. Then let $t$ be the vertex of $P$ contacting $H$. Then, since $P(H) \setminus \{z\} = \emptyset$, path $P$ has at most three vertices not in $H$, i.e. say $t,t_1,v$. Then, since $dist(za_2,v) \geq 7$, $t$ is nonadjacent to any vertex of $H$ except $c_2$. But then $d,a_1,a_2,b_1,b_2,c_1,c_2,t$ (with center $d$) induce an $S_{2,2,3}$, which is a contradiction.

The assertion concerning $dist(a_1a_2,v) \leq 6$ can be proved similarly.

\medskip

$(iii)$: This follows by statement $(ii)$ and by Lemma \ref{theo:N7} (with $G[V(K) \cup \{v\}]$ instead of $G$ and with $za_1$ and $a_1a_2$ instead of $xy$).

\medskip

$(iv)$: If $K'$ is a component of $G[V \setminus \{z\}]$, $K' \neq K$, $K' \neq K_{xy}$, and $K'$ contains an edge then, since $z$ is adjacent to the $S_{2,2,2}$ $H$, this leads to an $S_{2,2,3}$ with center $z$. If $K'$ contains only one vertex, say $z'$, then, since $z,z',a_1,a_2$ induce a paw and $d_G(z')=1$, it follows by Observation \ref{triangleneighbdeg1} that $z'$ is white.
\qed

\begin{lemma}\label{P(H)nonemptyimplCase5}
If $P(H) \setminus \{z\} \neq \emptyset$ then $d$ is white and $a_1a_2 \in M$ is forced.
\end{lemma}

{\bf Proof.} Recall that, by Proposition \ref{P5}, $z$ is adjacent to $a_1,a_2$ and in this case, to $b_1$. Since $P(H) \setminus \{z\} \neq \emptyset$, let $t \in P(H) \setminus \{z\}$. Then by definition of $P(H)$, let $t,t_2,t_3,t_4$ induce a $P_4$ with edges $tt_2,t_2t_3,t_3t_4$ such that $t_2,t_3,t_4$ do not contact $H$. Note that, by definition of $z$, vertex $t$ is nonadjacent to $z_3,z_4,z_5$, and since $d_G(a_2)=2$, $ta_2 \notin E$. Thus, only $t \join \{b_1,b_2\}$ or $t \join \{c_1,c_2\}$ is possible. For proving Lemma \ref{P(H)nonemptyimplCase5}, let us consider the following two cases which are exhaustive by Proposition \ref{prop:adjacencyH}.

\medskip

{\bf Case 1.} $t \join \{b_1,b_2\}$.

\medskip

We first claim:
\begin{equation}\label{b1b2tnonadjzz2}
tz \notin E \mbox{ and } tz_2 \notin E.
\end{equation}

{\em Proof.}
Since $G$ is diamond-free and thus, $b_1,b_2,t,z$ do not induce a diamond, we have $tz \notin E$.
Recall that $t \cojoin \{a_2,z_3,z_4,z_5\}$.
Since $z_2,z,a_2,t,b_2,z_3,z_4,z_5$ (with center $z_2$) do not induce an $S_{2,2,3}$, we have $tz_2 \notin E$.
$\diamond$

\medskip

Next we claim:
\begin{equation}\label{b1b2tnonadja1}
ta_1 \notin E.
\end{equation}

{\em Proof.} Clearly, if $ta_1 \in E$ then, since $t \join \{b_1,b_2\}$, $tc_1 \notin E$ and $tc_2 \notin E$. By (\ref{b1b2tnonadjzz2}), $tz_2 \notin E$ but then $a_1,t,b_2,z,z_2,d,c_1,c_2$ (with center $a_1$) induce an $S_{2,2,3}$, which is a contradiction. Thus $ta_1 \notin E$.
$\diamond$

\medskip

Next we claim:
\begin{equation}\label{b1b2tnonadjc1}
tc_1 \notin E.
\end{equation}

{\em Proof.} Suppose to the contrary that $tc_1 \in E$. Recall that by (\ref{b1b2tnonadjzz2}), $tz_2 \notin E$. Thus clearly $t_2 \neq z_2$.
First we claim that $zt_2 \notin E$: Otherwise, if $zt_2 \in E$ and since $G$ is butterfly-free, we have $z_2t_2 \notin E$ but then $z_3t_2 \notin E$ since otherwise,
$t_2,z,a_1,z_3,z_4,t,c_1,c_2$ (with center $t_2$) would induce an $S_{2,2,3}$ (vertex $t_2$ is nonadjacent to $z_4$ by definition of $z$).
Then $z,a_1,d,t_2,t,z_2,z_3,z_4$ (with center $z$) induce an $S_{2,2,3}$, which is a contradiction.
Thus, $zt_2 \notin E$.

\medskip

Next we claim that $zt_3 \notin E$: If $zt_3 \in E$ then $z,t_3,t_2,t,b_1$ and $a_1,a_2,b_2$ induce a $G_2$, which is impossible by Assumption 4.

\medskip

Thus $zt_3 \notin E$ but now, $t,t_2,t_3,c_1,c_2,b_1,z,a_1$ (with center $t$) induce an $S_{2,2,3}$, which is a contradiction. Thus, (\ref{b1b2tnonadjc1}) is shown.
$\diamond$

\medskip

Since $t$ is nonadjacent to $a_1,a_2,c_1$, we have $tc_2 \in E$. Then $t,b_1,d,c_1,c_2$ induce a $C_5$. Finally we claim:
\begin{equation}\label{b1b2tadjc2}
a_1a_2 \in M \mbox{ is forced and } d \mbox{ is white}.
\end{equation}

{\em Proof.} As before, by (\ref{b1b2tnonadjzz2}), $tz_2 \notin E$ and thus, $t_2 \neq z_2$. Suppose that $a_1a_2 \notin M$, i.e., $za_2 \in M$. Then $a_1,z_2$ and $b_1$ are white, $d$ and $b_2$ are black and thus, $b_2t \in M$, and by the $C_5$ property, $dc_1 \in M$. Moreover, $t_2$ is white and thus, $t_3$ is black which implies that $z t_3 \notin E$.

Since $b_1,z,a_2,d,c_1,t,t_2,t_3$ (with center $b_1$) do not induce an $S_{2,2,3}$, we have $zt_2 \in E$.
Since $t_2,z_2$ are white, we have $t_2z_2 \notin E$.
Since $z,t_2,t,z_2,z_3,a_1,d,c_1$ (with center $z$)  do not induce an $S_{2,2,3}$, we have $t_2z_3 \in E$.

Note that $t_2z_4 \notin E$ by definition of $z$. But now, $t_2,z,a_2,z_3,z_4,t,c_2,c_1$ (with center $t_2$) induce an $S_{2,2,3}$, which is a contradiction.

Thus, $a_1a_2 \in M$ is forced, $d$ is white and (\ref{b1b2tadjc2}) is shown.
$\diamond$

\medskip

{\bf Case 2.} $t \join \{c_1,c_2\}$.

\medskip

Then clearly, since $G$ is diamond-free, $td \notin E$, since $d_G(a_2)=2$, $ta_2 \notin E$, and $t$ is nonadjacent to $b_1$ or $b_2$.

First we claim:
\begin{equation}\label{c1c2tnonadjz}
tz \notin E.
\end{equation}

{\em Proof.} If $tz \in E$ then, since $G$ is diamond-free, $ta_1 \notin E$. Then, since $G$ is butterfly-free, $tb_1 \notin E$. But now,
$t,c_1,d,b_1,z$ and $a_1,c_2$ induce a $G_3$, which is impossible by Assumption~5. Thus, (\ref{c1c2tnonadjz}) is shown.
$\diamond$

\medskip

Note that $tz_2 \notin E$ since $z_2,z,a_2,t,c_1,z_3,z_4,z_5,$ (with center $z_2$) do not induce an $S_{2,2,3}$.

We claim:
\begin{equation}\label{c1c2tnonadja1a2}
ta_1 \notin E.
\end{equation}

{\em Proof.} If $ta_1 \in E$ then $tb_1 \notin E$ and $tb_2 \notin E$. Recall that $tz \notin E$ and
$tz_2 \notin E$. But then $z,z_2,z_3,b_1,b_2,a_1,t,c_2$ (with center $z$) induce an $S_{2,2,3}$, which is a contradiction. Thus $ta_1 \notin E$.
$\diamond$

\medskip

Since $ta_1 \notin E$ and $ta_2 \notin E$, it follows that $tb_1 \in E$ or $tb_2 \in E$.
Recall that $za_1 \notin M$, i.e., only $za_2 \in M$ or $a_1a_2 \in M$ is possible. Finally we show:
\begin{equation}\label{c1c2tadjb1orb2}
a_1a_2 \in M \mbox{ is forced and } d \mbox{ is white}.
\end{equation}

{\em Proof.}
Suppose that $a_1a_2 \notin M$, i.e., $za_2 \in M$. Then $a_1$ and $b_1$ are white, which implies that $d$ and $b_2$ are black.

\medskip

First let $tb_2 \in E$. Since $d$ is black, by the $C_3$ property with respect to triangle $c_1c_2t$, we have $c_2t \in M$, which is a contradiction since $b_2$ is black.

\medskip

Now let $tb_1 \in E$. Then, since $b_1$ is white, $t$ is black. Then, since $d$ is black, $c_1$ is white and thus, $tc_2 \in M$.

Recall that $tz \notin E$ and $tz_2 \notin E$. Let $d'$ be an $M$-mate of $d$ such that $dd' \in M$, and let $b'_2$ be an $M$-mate of $b_2$ such that
$b_2b'_2 \in M$. Clearly, $d'z \notin E$ and $b'_2z \notin E$.

\medskip

We claim that $d'z_2 \notin E$ and $b'_2z_2 \notin E$: If $b'_2z_2 \in E$ then $z_2,b'_2,b_2,z,a_2,z_3,z_4,z_5$ (with center $z_2$) would induce an $S_{2,2,3}$, and analogously, if $d'z_2 \in E$ then $z_2,d',d,z,a_2,z_3,z_4,z_5$ (with center $z_2$) would induce an $S_{2,2,3}$. Thus, $d'z_2 \notin E$ and $b'_2z_2 \notin E$.

\medskip

Furthermore, since $G$ is butterfly-free, we have $b_1d' \notin E$ or $b_1b'_2 \notin E$.
If $b_1d' \notin E$ then $b_1,t,c_2,d,d',z,z_2,z_3$ (with center $b_1$) would induce an $S_{2,2,3}$ (note that $z_3d' \notin E$), and
if $b_1b'_2 \notin E$ then $b_1,t,c_2,b_2,b'_2,z,z_2,z_3$ (with center $b_1$) would induce an $S_{2,2,3}$
(note that $z_3b'_2 \notin E$), which is a contradiction.

\medskip

Thus, (\ref{c1c2tadjb1orb2}) is shown.
$\diamond$

\medskip

This completes the proof of Lemma \ref{P(H)nonemptyimplCase5}.
\qed

\medskip

Then let us summarize the results for the case $zb_1 \in E$ as follows. Recall that $P(H)$ can be computed in polynomial time.

\begin{lemma}\label{lemm:zb1}
Assume that $zb_1 \in E$.
\begin{itemize}
\item[$(i)$] If $P(H) \setminus \{z\} \neq \emptyset$ then by Lemma $\ref{P(H)nonemptyimplCase5}$, $a_1a_2 \in M$ and $d$ is a white vertex, that is, one can easily detect a white vertex of $H$.
\item[$(ii)$] If $P(H) \setminus \{z\} = \emptyset$, then $z$ is a cut-vertex of $G$ and in particular, if $K$ denotes the component of $G[V \setminus \{z\}]$ containing $H$, one can check in polynomial time $-$ by Lemma \ref{theo:N7} $-$ whether $G[K \cup \{z\}]$ admits a d.i.m.\ containing $za_2$ and a d.i.m.\ containing $a_1a_2$.
\end{itemize}
\end{lemma}

\subsection{Deleting $S_{2,2,2}$'s in $G[Y]$}

Let $H$ be an induced $S_{2,2,2}$ in $G[Y]$ with vertices $V(H)=\{d,a_1,a_2,b_1,b_2,c_1,c_2\}$ and edges $da_1$, $db_1$, $dc_1$, $a_1a_2$, $b_1b_2$, $c_1c_2$ (as above). Let $p := \min \{i : N_i \cap V(H) \neq \emptyset\}$, that is, $N_p$ is the $xy$-distance level with smallest distance to $xy$ to which a vertex of $H$ belongs (in particular $p \geq 4$ by construction).

\medskip

We say that $H$ is {\em critical} if
\begin{itemize}
\item[$(i)$] there is a contacting vertex for $H$, say $z$, with $z \in N_{p-1}$ and $p \geq 6$, such that $za_1,za_2,zb_1 \in E$, and
\item[$(ii)$] $P(H) = \{z\}$ (cf. Section \ref{section5.2}).
\end{itemize}

Otherwise, $H$ is {\em non-critical}.

\medskip

Recall that if $H$ is a critical $S_{2,2,2}$, then either $za_2 \in M$ or $a_1a_2 \in M$, so that $G[Y]$ has a d.i.m.\ $M$ only if $za_2 \in M$ or $a_1a_2 \in M$,
and by Proposition \ref{prop:P(H)second}, we can assume that $G-z$ has only two components.

\medskip

Then let us consider the following procedure deleting a critical $S_{2,2,2}$, which is correct and can be executed in polynomial time by Lemma
\ref{lemm:zb1} $(ii)$ and by the above (and in particular, by the distance properties of $K$ - see Proposition \ref{prop:P(H)second} $(ii)$).

\begin{proc}[Delete-Critical-$S_{2,2,2}$]\label{criticalS222}

\begin{tabbing}	
xxxxxxx \= \kill\\
{\bf Input:} \> Subgraph $G[Y]$ and a critical $S_{2,2,2}$, say $H$.\\

{\bf Task:} \> Return either a proof that $G[Y]$ has no d.i.m., or a subgraph of $G[Y]$, say $G'$\\
\> such that\\
\> $(i)$ $G'$ does not contain some vertices of $H$, and\\
\> $(ii)$ $G'$ has a d.i.m.\ if and only if $G[Y]$ has a d.i.m.
\end{tabbing}

\begin{itemize}
  \item [$(a)$] Compute the component, say $K$, of $G[V \setminus \{z\}]$ containing $H$. Then check whether $G[V(K) \cup \{z\}]$ admits a d.i.m. containing $za_2$ and a d.i.m. containing $a_1a_2$.
  \item [$(b)$] Consider the following exhaustive occurrences:
\begin{itemize}
  \item[$(b.1)$] {\bf if} $G[V(K) \cup \{z\}]$ admits no d.i.m.\ containing $za_2$ and no d.i.m.\ containing $a_1a_2$ {\bf then} return ``$G[Y]$ has no d.i.m.''

  \item[$(b.2)$] {\bf if} $G[V(K) \cup \{z\}]$ admits a d.i.m.\ containing $za_2$ and no d.i.m.\ containing $a_1a_2$ {\bf then} in $G[Y]$: $(i)$ delete all vertices of $V(K) \setminus \{a_1,a_2\}$; $(ii)$ color $z,a_2$ black and color $a_1$ white. Then let $G'$ be the resulting graph.

  \item[$(b.3)$] {\bf if} $G[V(K) \cup \{z\}]$ admits no d.i.m.\ containing $za_2$ and a d.i.m.\ containing $a_1a_2$ {\bf then} in $G[Y]$: $(i)$ delete all vertices of $V(K) \setminus \{a_1,a_2\}$; $(ii)$ color $a_1,a_2$ black and color $z$ white. Then let $G'$ be the resulting graph.

  \item[$(b.4)$] {\bf if} $G[V(K) \cup \{z\}]$ admits a d.i.m.\ containing $za_1$ and a d.i.m.\ containing $a_1a_2$ {\bf then} in $G[Y]$, delete all vertices of $V(K) \setminus \{a_1,a_2\}$. Then let $G'$ be the resulting graph.
\end{itemize}
  \item [$(c)$] {\bf if} a contradiction arises in the vertex coloring {\bf then} STOP and return ``$G[Y]$ has no d.i.m.'' {\bf else} STOP and return $G'$.
\end{itemize}
\end{proc}

\begin{lemma}\label{lemm:final}
For any non-critical $S_{2,2,2}$ $H$, one can detect a white vertex of $H$ or a peripheral triangle with a vertex in $H$ in polynomial time.
\end{lemma}

{\bf Proof.} 
Lemma \ref{lemm:final} follows by Proposition \ref{obse:p5} and by Lemmas \ref{lemm:zb2} and \ref{lemm:zb1}.
\qed \\

Then let us consider the following procedure deleting $S_{2,2,2}$'s in $G[Y]$, which is correct and can be executed in polynomial time by Lemma \ref{lemm:final} $-$ which involves Lemma \ref{theo:N7} $-$ and since the solution algorithm of Hertz et al. (see Theorem \ref{DIMpolresults} $(iii)$) can be executed in polynomial time.

\begin{proc}[Delete-$S_{2,2,2}$]\label{delS222}

\begin{tabbing}	
xxxxxxxxx \= \kill\\
{\bf Input:} \> Graph $G[Y]$ with a partial coloring of its vertices.\\
{\bf Output:} \> A d.i.m.\ of $G[Y]$ $($consistent with the partial coloring$)$ or\\
 \> a proof that $G$ has no d.i.m.\ $($consistent with the partial coloring$)$.
\end{tabbing}
\begin{tabbing}
xxxxx\=xx\=xx\=xx\=xx\=xx\=xx \kill
{\bf begin}\\
$(a)$ \>\> Set $F:=G[Y]$;\\
$(b)$ \>\> {\bf while} $F$ contains a critical $S_{2,2,2}$, say $H$ {\bf do} \\
\>\>\> {\bf begin}\\
\>\>\>\> Apply Procedure $\ref{criticalS222}$ to $H$; \\
\>\>\>\> {\bf if} it returns that $G[Y]$ has no d.i.m.\ {\bf then} return ``$G[Y]$ has no d.i.m''\\
\>\>\>\> {\bf else} let $F'$ be the resulting subgraph of $F$: then set $F:= F'$; \\
\>\>\> {\bf end};\\
$(c)$ \>\> {\bf while} $F$ contains a non-critical $S_{2,2,2}$, say $H$ {\bf do} \\
\>\>\> {\bf begin}\\
\>\>\>\> either detect a white vertex $h \in V(H)$ and apply the Vertex Reduction to $h$\\
\>\>\>\> or detect a peripheral triangle of $G$, say $abc$ involving some vertex of $H$,\\
\>\>\>\> and apply the Peripheral Triangle Reduction to $abc$; \\
\>\>\>\> {\bf if} a contradiction arises in the black-white vertex coloring\\
\>\>\>\> {\bf then} STOP with failure\\
\>\>\>\> {\bf else} let $F'$ be the resulting subgraph of $F$: then set $F:= F'$; \\
\>\>\> {\bf end};\\
$(d)$ \>\> Apply the solution algorithm of Hertz et al. $($see Theorem $\ref{DIMpolresults}$ $(iii))$ \\
\>\> to determine if $F$ $\{$which is $S_{2,2,2}$-free by the above$\}$ has a d.i.m.: \\
\>\> {\bf if} $F$ has a d.i.m.\ {\bf then} STOP and return such a d.i.m.;\\
\>\> {\bf else} STOP and return ``$G[Y]$ has no d.i.m.'';\\
{\bf end}.
\end{tabbing}
\end{proc}

\section{A polynomial algorithm for DIM on $S_{2,2,3}$-free graphs}

The following procedure is part of the algorithm:

\medskip

\begin{proc}[DIM-with-$xy$]\label{DIMwithxy}

\begin{tabbing}	
xxxxxxx \= \kill\\
{\bf Input:} \> A connected $S_{2,2,3}$-free graph $G = (V,E)$, which satisfies\\
\> Assumptions $1-6$ of Section $\ref{exclforced}$ and\\
\> an edge $xy \in E$ which is part of an induced $P_3$ in $G$.\\
{\bf Task:} \> Return a d.i.m.\ $M$ with $xy \in M$ $($STOP with success$)$ or\\
\> a proof that $G$ has no d.i.m.\ $M$ with $xy \in M$ $($STOP with failure$)$.
\end{tabbing}

\begin{itemize}

\item[$(a)$] Set $M:= \{xy\}$. Determine the distance levels $N_i = N_i(xy)$ 
    with respect to $xy$.

\item[$(b)$] Check whether $N_1$ is an independent set $($see fact $(\ref{N1subI}))$ and $N_2$ is the disjoint union of edges and isolated vertices $($see fact $(\ref{N2M2S2}))$. If not, then STOP with failure.

\item[$(c)$]  For the set $M_2$ of edges in $N_2$, apply the Edge Reduction for every edge in $M_2$. Moreover, apply the Edge Reduction for each edge $bc$ according to fact $(\ref{triangleaN3bcN4}$) and then for each edge $u_it_i$ according to Observation $\ref{lemm:structure2}$ $(v)$.

\item[$(d)$]  {\bf if} $N_4 = \emptyset$ then apply the approach described in Section $\ref{ColoringG[X]}$. Then either return that $G$ has no d.i.m.\ $M$ with $xy \in M$ or return $M$ as a d.i.m.\ with $xy \in M$.

\item[$(e)$]  {\bf if} $N_4 \neq \emptyset$ {\bf then} for $X := \{x,y\} \cup N_1 \cup N_2 \cup N_3$ and $Y := V \setminus X$ $($according to the results of Sections $\ref{ColoringG[Y]}$ and $\ref{sec:N7nonempty})$ {\bf do}

\begin{itemize}
\item[$(e.1)$] Compute all black-white $xy$-colorings of $G[X]$. If no such feasible $xy$-coloring exists then STOP with failure.

\item[$(e.2)$]  {\bf for each} $xy$-coloring of $G[X]$  {\bf do}
\begin{enumerate}
\item[$(e.2.1)$]  Derive a partial coloring of $G[Y]$ by the forcing rules;
\item[ ]{\bf if} a coloring contradiction arises {\bf then} STOP with failure.

\item[$(e.2.2)$] Apply Procedure $\ref{delS222}$; {\bf if} it returns a d.i.m.\ of $G[Y]$ {\bf then} STOP with success and return the $xy$-coloring of $G$ derived by the $xy$-coloring of $G[X]$ and by such a d.i.m.\ of $G[Y]$.
\end{enumerate}

\item[$(e.3)$] STOP with failure.
\end{itemize}
\end{itemize}

\end{proc}

\begin{theorem}\label{theo:procedureDIMxy}
Procedure $\ref{DIMwithxy}$ is correct and runs in polynomial time.
\end{theorem}

{\bf Proof.} The correctness of the procedure follows from the structural analysis of $S_{2,2,3}$-free graphs with a d.i.m.

\medskip

The polynomial time bound follows from the fact that Steps (a) and (b) can clearly be done in polynomial time, Step (c) can be done in polynomial time since the Edge Reduction can be done in polynomial time, and Steps (d) and (e) can be done in polynomial time by the results in Sections $\ref{ColoringG[X]}$, $\ref{ColoringG[Y]}$,
 and \ref{sec:N7nonempty}.
\qed

\begin{algo}[DIM-$S_{2,2,3}$-free]\label{DIMS223fr}

\begin{tabbing}	
xxxxxxx \= \kill\\
{\bf Input:} \> A connected $S_{2,2,3}$-free graph $G = (V,E)$, which satisfies\\
\> Assumptions $1-6$ of Section $\ref{exclforced}$.\\

{\bf Task:} \> Determine a d.i.m.\ of $G$ if there is one, or find out that $G$ has no d.i.m.
\end{tabbing}

\begin{itemize}
\item[$(A)$] Check whether $G$ has a single edge $uv \in E$ which is a d.i.m.\ of $G$. If yes then select such an edge as output and STOP - this is a d.i.m.\ of $G$.
$\{$Otherwise, every d.i.m.\ of $G$ would have at least two edges.$\}$

\item[$(B)$] {\bf for each} edge $xy \in E$ in a $P_3$ of $G$, carry out Procedure $\ref{DIMwithxy}$;\\
{\bf if} it returns ``STOP with failure'' for all edges $xy$ in a $P_3$ of $G$ {\bf then} STOP -- $G$ has no d.i.m.\ {\bf else} STOP and return a d.i.m.\ of $G$.
\end{itemize}

\end{algo}

\begin{theorem}\label{theo:procedureDIM}
Algorithm $\ref{DIMS223fr}$ is correct and runs in polynomial time. Thus, DIM can be solved in polynomial time for $S_{2,2,3}$-free graphs.
\end{theorem}

{\bf Proof.} The correctness of the procedure follows from the structural analysis of $S_{2,2,3}$-free graphs with a d.i.m.\ In particular, Step (A) is obviously correct, and for Step (B), recall Observation \ref{obse:xy-in-P3}.

\medskip

The time bound follows from the fact that Step (A) can be done in polynomial time, and Step (B) can be done in polynomial time by Theorem \ref{theo:procedureDIMxy}.
\qed

\medskip

{\bf Acknowledgment.}
We gratefully thank the anonymous reviewers for their comments and corrections.
The second author would like to witness that he just tries to pray a lot and is not able to do anything without that - ad laudem Domini.

\begin{footnotesize}

\end{footnotesize}

\end{document}